\documentclass[11pt]{article}
\usepackage{graphicx,CJK}
\usepackage{amsmath,amssymb,amsthm}
\usepackage{amsfonts,mathrsfs}
\usepackage{bm,indentfirst,color}
\usepackage{caption,subcaption}

\allowdisplaybreaks[4]

\begin{document}
\renewcommand{\d}{\mathrm{d}}
\newtheorem{theorem}{Theorem}[section]
\numberwithin{equation}{section}

\renewcommand{\figurename}{Fig.}

\title{Onsager-theory-based dynamic model for nematic phases of bent-core molecules and star molecules}
\author{Jie Xu$^1$ and Pingwen Zhang$^2$\vspace{12pt}\\
\small
$^1$Department of Mathematics, Purdue University, West Lafayette 47907, USA
\\
\small
$^2$LMAM \& School of Mathematical Sciences, Peking University, Beijing 100871, China\\
\small
Email: xu924@purdue.edu, pzhang@pku.edu.cn
}
\date{\today}
\maketitle

\begin{abstract}
We construct the molecular model and the tensor model for the dynamics of the nematic phases of bent-core molecules and star molecules in incompressible fluid. 
We start from the molecular interaction and the molecule--fluid friction, 
and write down a general formulation based on the molecular shape and the free energy. 
Then we incorporate an Onsager-theory-based static tensor model that is determined by the molecular shape. 
In this way, the terms in the molecular model are fully determined by the molecular shape and expressed by physical parameters. 
For bent-core molecules and star molecules, the model shares the same form, but is different in the coefficients. 
With the polar interaction and elastic energy taken into account, and the convection and diffusion included both spatially and orientationally, the model is suitable for inhomogeneous flows. 
We adopt the quasi-equilibrium approximation to derive the tensor model with energy dissipation maintained. 
Numerical simulation is carried out focusing on the shear flow problem using both the molecular and the tensor models. 
We choose the parameters near the transition region of different equilibrium nematic phases and examine the effect of molecular shape on the flow modes. 
The tensor model proves to exhibit all the flow modes found in the molecular model. 

\vspace{12pt}
\end{abstract}

\section{Introduction}
The liquid crystalline flows are studied extensively in the last few decades. 
The majority of works focus rod-like molecules that can exhibit uniaxial nematic phase in equilibrium. 
The earliest and simplest approach is the macroscopic Ericksen-Leslie theory \cite{Ericksen_Leslie}. 
However, this approach is insufficient to investigate singular phenomena, such as defects. 
For the microscopic approach, Doi \cite{Doi_book} established the kinetic equation of the density function (the Smoluchowski equation), which we call the molecular model. 
Doi theory has been applied to study the spatially homogeneous shear flow problem for rod-like molecules \cite{Dyn4,Dyn3,Dyn1,Dyn2}. 
%The predictions made by the theory proved to be exact. 
It has also been extended to inhomogeneous flows \cite{DynInh2,Marrucci_JNNFM1992,WangELiuZhang_PRE2002,Yu1,DynInh1}. 
Despite its great success, the simulation is time-consuming, making its application to inhomogeneous flows rather restricted. 
To reduce the dimension of variables, many works aim to construct models in which the orientation is described by tensors. 
%Models of this kind are called tensor models. 
Apart from some phenomenological tensor models \cite{Beris_book,pre1998}, most tensor models are obtained by closure approximation of Doi theory \cite{advani1987use,advani1990closure,clos3,feng1998closure,clos1,Ilg2,clos2,wang1997comparative,Yu2}. 
With various closure approximations for different types of flows, the tensor models have proved to be able to capture the phenomena in the molecular model, although no one closure approximation can recover all the phenomena. 
The Doi theory, the tensor models and the Ericksen-Leslie theory have been shown closely bonded \cite{RodModel}. 

When the molecule is not axisymmetric, it is possible to exhibit multiple nematic phases. 
As a representative, bent-core molecules have attracted much attention. 
Besides the uniaxial nematic phase, they have proved to be able to show the biaxial nematic phase \cite{BiExp_prl2004_2,BiExp_prl2004}, and the twist-bend phase \cite{Ntb,Ntb2,prl_111_067801}, a modulated nematic phase with polar order. 
The phase behavior of bent-core molecules has also been discussed theoretically \cite{pre73,PRL_115_147805,prl_3D_2014,pre2013,pre2014_2}. 
The dynamics of bent-core molecules is expected to be much more complicated and fascinating. 
However, the works on the dynamics of non-axisymmetric liquid crystals are sparse. 
%The construction of model also follows two distinct routes. 
%One approach is to discuss the dynamics of a certain nemaatic phase, which is done 
Macroscopic dynamic models have been proposed for the biaxial nematic phase \cite{EC_pra2,EC_pra1,EC_pra3} as an extension of the Ericksen-Leslie theory. 
But their applications are even more limited than the Ericksen-Leslie theory since multiple nematic phases may coexist. 
%The other approach is to start from microscopic interaction. 
As for the microscopic approaches, the Smoluchowski equation is adopted to investigate the dynamics of ellipsoids \cite{pre78,jr2009} and bent-core molecules \cite{cms2010}. 
These works focus on the homogeneous shear flow problem and have obtained flow modes different from rod-like molecules. 
Nevertheless, since the equilibrium theory of bent-core molecules was far from well-established at the time these works were carried out, 
the model in these works includes only the local biaxial interaction. 
This makes the model only specifically suitable for the homogeneous flows, because the polar interaction and modulation have proved to be cruicial in inhomogeneous systems. 
In addition, the terms in their model are derived from different molecular architectures. 
This inconsistency blurs the effect of molecular shape on the flow features, which would be the most interesting problem to be studied. 

In a recent work \cite{BentModel}, we construct a static tensor model from the Onsager theory for the nematic phases of bent-core molecules and star molecules (see Fig. \ref{mol}). 
We assume that the molecule is rigid and consists of spheres. 
The free energy is a functional of three tensors, one first-order (vector) and two second-order symmetric, with gradient terms included, enabling the free energy to describe polarity and modulation. 
The form of the free energy is determined by molecular symmetry, and
the coefficient of each term is derived as a function of physical parameters. 
In this work, we build the free energy into the dynamic model. 
To incorporate the molecule-fluid interaction, we adopt the same molecular architecture in the static model and consider the friction between the fluid and spheres. 
In this way, we are able to write down the molecular model, with the energy dissipation law, fully based on molecular architecture and physical parameters. 
Similar to the static model, for bent-core molecules and star molecules, the model shares the same form, but different in coefficients. 
The model includes convection terms that originate from the molecular-fluid interaction, and diffusion terms that originate from molecular interaction, both spatially and orientationally. 
Together with the polarity and modulation, the model is applicable to inhomogeneous flows. 
We then write down the tensor model by deriving the equation of the three tensors appearing in the free energy from the Smoluchowski equation. 
The high-order tensors appearing in the tensor model are expressed by the three tensors using quasi-equilibrium approximation, a generalization of the Bingham closure. 
When adopting the quasi-equilibrium approximation, the tensor model retains the energy dissipation law. 

For the numerical simulation, we restrained our attention to the shear flow problem. 
In particular, we choose the parameters in the vicinity of the uniaxial-biaxial phase boundary, which is not studied previously. 
We focus on adjusting the parameters describing the molecular shape, and investigate how the flow modes are altered. 
%Using both the molecular model and the tensor model, we examine the flow mode sequences as the shear rate increases, with varying bending angles that exhibit different nematic phases. {\color{red}(Emphasize shape: star molecules). We focus on the parameters in the transition region of uniaxial and biaxial nematic phases. We examine the effect of molecular shape on the flow modes. }
Also, we compare the results from the molecular model and the tensor model. 
The tensor model is able to exhibit all the flow modes found in the molecular model, although under different parameters. 

The paper is organized as follows. 
In Sec. \ref{molecularmodel} we derive the molecular model. In Sec. \ref{tensormodel} we derive the tensor model and prove the energy dissipation along with the quasi-equilibrium closure approximation. In Sec. \ref{results} we use both molecular model and tensor model to examine the shear flow problem. A conclusion is drawn in Sec. \ref{concl}. 

\section{Molecular model\label{molecularmodel}}
\subsection{Notations}
We view the molecules that form liquid crystalline states as fully rigid. 
Thus, we may choose a body-fixed orthogonal frame $(\hat{O};\bm{m}_1,\bm{m}_2,\bm{m}_3)$ 
to describe the position and the orientation of a molecule. 
In a space-fixed orthogonal coordinate system $(O;\bm{e}_1,\bm{e}_2,\bm{e}_3)$, they can be expressed in terms of $\bm{x}=\overrightarrow{O\hat{O}}$ 
and a three-dimensional proper rotation $P\in SO_3$. 
In the language of matrix, $P$ is a $3\times 3$ orthogonal with det$P=1$ such that 
\begin{equation}\label{RotP}
\left(
  \bm{m}_1,
  \bm{m}_2,
  \bm{m}_3
\right)
=
\left(
  \bm{e}_1,
  \bm{e}_2,
  \bm{e}_3
\right)P.
\end{equation}
The elements of $P^T=(m_{ij})$ are the components of $\bm{m}_i$, denoted by 
$$
m_{ij}=\bm{m}_i\cdot\bm{e}_j.
$$
In some cases, we need to specify a point on the molecule, and we use its coordinates 
$\bm{\hat{r}}$ in the body-fixed frame $(\hat{O};\bm{m}_1,\bm{m}_2,\bm{m}_3)$. 
Every $P\in SO_3$ can be expressed by Euler angles $\alpha, \beta, \gamma$: 
\begin{align}
&P(\alpha,\beta,\gamma)\nonumber\\
=&\left(
\begin{array}{ccc}
 \cos\alpha &\quad -\sin\alpha\cos\gamma &\quad\sin\alpha\sin\gamma\\
 \sin\alpha\cos\beta &\quad\cos\alpha\cos\beta\cos\gamma-\sin\beta\sin\gamma &
 \quad -\cos\alpha\cos\beta\sin\gamma-\sin\beta\cos\gamma\\
 \sin\alpha\sin\beta &\quad\cos\alpha\sin\beta\cos\gamma+\cos\beta\sin\gamma &
 \quad -\cos\alpha\sin\beta\sin\gamma+\cos\beta\cos\gamma
\end{array}
\right),\label{EulerRep}
\end{align}
with
$$
\alpha\in [0,\pi],\ \beta,\gamma\in [0,2\pi). 
$$
The uniform probability measure on $SO_3$ is given by
$$
\d\nu=\frac{1}{8\pi^2}\sin\alpha\d\alpha\d\beta\d\gamma. 
$$

For the notations of tensors, we use the summation over repeated indices. 
The product of several tensors without operators is recognized as tensor product: $\bm{m}_1\bm{m}_2\bm{m}_3$ represents a third order tensor with the $(i,j,k)$ component $m_{1i}m_{2j}m_{3k}$. 
If a tensor contraction involves first or second order tensor, we also use the single and double dots: suppose we have a first order tensor $\bm{p}$, a second order tensor $Q$, and a fourth order tensor $R$, then 
\begin{align}
  (Q\cdot\bm{p})_i=Q_{ij}p_j,\quad (\bm{p}\cdot R)_{jkl}=p_iR_{ijkl},\quad (Q:R)_{kl}=Q_{ij}R_{ijkl}. 
\end{align}
%{\color{red}(Define notations for tensors)}

To describe the number of molecules with certain position $\bm{x}$ and orientation $P$, we introduce the density function $f(\bm{x},P)$. Moreover, we split $f(\bm{x},P)$ into the local concentration $c(\bm{x})$ and the orientational distribution $\rho(\bm{x},P)$, 
\begin{equation}
  c(\bm{x})=\int\d\nu f(\bm{x},P), \qquad \rho(\bm{x},P)=f(\bm{x},P)/c(\bm{x}). 
\end{equation}
The notation $\left<\cdot\right>$ represents the average about $\rho(\bm{x},P)$, 
$$
\left<(\cdot)\right>=\int\d\nu (\cdot)\rho(\bm{x},P). 
$$

The differential operators on $SO(3)$ are involved when discussing the motion of the rigid molecules. 
In specific, we use $L_i$ to denote the derivatives along the infinitesimal rotation about $\bm{m}_i$. 
The operators $L_i$ can be expressed by derivatives of Euler angles, 
\begin{align}
L_1&=\frac{\partial}{\partial\gamma},\\
L_2&=\frac{-\cos\gamma}{\sin\alpha}\left(\frac{\partial}{\partial\beta}
-\cos\alpha\frac{\partial}{\partial\gamma}\right)
+\sin\gamma\frac{\partial}{\partial\alpha},\\
L_3&=\frac{\sin\gamma}{\sin\alpha}\left(\frac{\partial}{\partial\beta}
-\cos\alpha\frac{\partial}{\partial\gamma}\right)
+\cos\gamma\frac{\partial}{\partial\alpha}. \label{DiffRep}
\end{align}
%The above expressions can be found in \cite{}. 
Denote $L=(L_1,L_2,L_3)^T$. If a vector-valued function $\bm{a}(P)$ is expressed as $\bm{a}(P)=a_1(P)\bm{m}_1+a_2(P)\bm{m}_2+a_3(P)\bm{m}_3$, then the divergence is defined by 
$$
L\cdot\bm{a}=L_1a_1+L_2a_2+L_3a_3.
$$
We may verify the following properties using the above definition. 
Acting the operators on $m_{ij}$, we have
\begin{align}
L_i\bm{m}_{j}&=\epsilon_{ijk}\bm{m}_{k}. \label{diffL}
\end{align}
Here we use the Levi-Civita symbol, 
$$
\epsilon_{k_1k_2k_3}=\left\{
\begin{array}{ll}
  1,&(k_1k_2k_3)=(123),(231),(312); \\
  -1,&(k_1k_2k_3)=(132),(213),(321); \\
  0,&\mbox{otherwise}. 
\end{array}
\right.
$$
The operators also satisfy the integration by parts on $SO(3)$, 
\begin{align}
\int\d\nu f(L_ig)&= -\int\d\nu (L_if)g. \label{IntPart}
\end{align}

%Notations for tensors. 

\subsection{General formulation}
The motion of rigid molecules includes translation and rotation, driven by molecular interaction and molecule--fluid interaction. 
In general, the translation and rotation can be coupled. 
But in what follows, we will deduce them separately under various approximations. 
To let our discussion be specific, we assume that a rigid molecule consists of spheres of the diameter $D$ and the mass $m_0$. 
In this case, the architecture of a molecule is given by the number density 
of the sphere centers $\hat{\rho}(\hat{\bm{r}})$ in the body fixed frame $(\hat{O};\bm{m}_1,\bm{m}_2,\bm{m}_3)$, and we assume that the center of mass is located at $\hat{O}$. 
The rigid molecules are dissolved in incompressible viscous fluid, and the molecule-fluid interaction stems from the friction between them. 
The frictional force between a sphere and the fluid is proportional to the relative velocity, given by $\bm{F}=-\zeta\bm{V}$, where $\zeta=3\pi D\eta_0$ is the friction constant. 

The molecular interaction induces a potential field $\mu(\bm{x},P)$ 
given by the functional derivative 
\begin{equation}
\mu=\frac{\delta F[f]}{\delta f}, 
\end{equation}
where $F[f]$ represents the free energy of a system with the number density $f(\bm{x},P)$ of rigid molecules. 
The free energy includes the contribution of the entropy and pairwise interaction, 
\begin{equation}
  F[f]=F_{entropy}[f]+F_r[f], \label{FreeEng0}
\end{equation}
where 
\begin{equation}
F_{entropy}=k_BT\int\d\bm{x}\d\nu~f\log f. \label{entropy}
\end{equation}
For bent-core molecules and star molecules, we will give the expression of $F_r$ later. 

\subsubsection{Smochulowski equation}
In general, the Smochulowski equation for the rigid molecules can be written as 
\begin{equation}
  \frac{\partial f}{\partial t}=-\nabla\cdot(f\bm{w})-L\cdot(f\bm{\omega}). 
\end{equation}
Here, for the molecule at the position $\bm{x}$ and the orientation $P$, we use $\bm{w}(\bm{x},P)$ to denote the velocity of the center of mass, and $\bm{\omega}(\bm{x},P)$ to denote the angular velocity. 
For both of them, we split the contribution of molecular interaction, $\bm{w}_m$, $\bm{\omega}_m$, and fluid--molecule interaction, $\bm{w}_f$, $\bm{g}$, by writing 
$$
\bm{w}=\bm{w}_m+\bm{w}_f,\quad \bm{\omega}=\bm{\omega}_m+\bm{g}. 
$$

We start from the rotation resulted from the molecular interaction. 
To derive this term, we assume that a molecule is rotating in the quiescent fluid, with the angular velocity $\bm{\omega}_m$ round the center of mass. 
The torque generated by the friction between a molecule and the quiescent fluid 
is the sum of frictional torque on each sphere, 
\begin{equation}
  -\bm{N}=\zeta\int \d\hat{\bm{r}} \hat{\rho}(\hat{\bm{r}})\hat{\bm{r}}\times(\bm{\omega}_m\times\hat{\bm{r}})=\frac{\zeta}{m_0}\bm{I}\bm{\omega}_m, \label{rot_work}
\end{equation}
where $\bm{I}$ is the moment of inertia of a molecule, calculated as 
\begin{equation}
  \bm{I}=m_0\int\d\hat{\bm{r}}\hat{\rho}(\hat{\bm{r}})\left(|\hat{\bm{r}}|^2
  -\hat{\bm{r}}\hat{\bm{r}}\right). \label{MoI}
\end{equation}
On the other hand, suppose the molecule is doing an infinitesimal rotation $\delta P=\delta t\bm{\phi}\times P$. 
Then the work done by the frictional torque is $-\bm{N}\cdot\bm{\phi}\delta t$, 
and shall equal to the variation of potential. Therefore, 
$$
\bm{N}\cdot\bm{\phi}\delta t=\mu(P+\delta P)-\mu(f)=\delta t\bm{\phi}\cdot L\mu, 
$$
yielding $\bm{N}=L\mu$. 
Comparing it with (\ref{rot_work}), we have 
$$
\bm{\omega}_m=-\frac{m_0}{\zeta}\bm{I}^{-1}L\mu.%\triangleq D_0\bm{I}^{-1}LV. 
$$
If we carefully choose $\bm{m}_i$ such that $\bm{I}$ is diagonal in the body-fixed frame, 
then we can write 
$$
\bm{\omega}_m=\sum_{i=1}^3\omega_i\bm{m}_i=-\sum_{i=1}^3(D_iL_i\mu)\bm{m}_i, 
$$
where the diffusion coefficients are given by 
\begin{equation}
D_i=\frac{m_0}{\zeta I_{ii}}. \label{rot_diff}
\end{equation}
Note that $\bm{\omega}_m$ actually gives a diffusion term $L\cdot(m_0\zeta^{-1}\bm{I}^{-1}fL\mu)$ in the Smoluchowski equation. 
Define 
\begin{equation}
  V=\frac{\delta F_r}{\delta f}, 
\end{equation}
then we can split $\mu$ as
\begin{equation}
  \mu=k_BT(\log f+1)+V. 
\end{equation}
and the diffusion term can also be written as 
\begin{equation}
  \sum_{i=1}^3D_i[k_BTL_i^2f+L_i(fL_iV)]. 
\end{equation}

Similarly, we derive $\bm{w}_m$ by considering the translation of a molecule in the quiescent fluid. 
Generally, $\bm{w}_m$ can be written as 
\begin{equation}
  \bm{w}_m=-\bm{J}(k_BT\nabla f+f\nabla V), 
\end{equation}
where the diffusion coefficient $\bm{J}$ is a $3\times 3$ positive matrix. 
To derive $\bm{J}$, we need to consider the hydrodynamic interaction, namely the interaction of different spheres through the fluid, which can be done using the Kirkwood theory (see \cite{Doi_book}). 
In Appendix, we will outline how to use the Kirkwood theory to calculate $\bm{J}$ and present the result for bent-core molecules. 
As a simple approximation, if we ignore the hydrodynamic interaction, then $\bm{J}$ will be a multiple of the identity matrix. 

Next we derive the translation and rotation generated by molecule--fluid interaction. 
Now we need to consider the motion of a molecule driven by the fluid with inhomogeneous velocity. 
We require that the velocity of the center of mass, $\bm{w}_f$, and the angular velocity, $\bm{g}$, minimize the frictional work. 
Denote by $\bm{u}(\hat{\bm{r}})$ and $\bm{u}_p(\hat{\bm{r}})$ the velocity of the fluid and the sphere at the point $\hat{\bm{r}}$% in the body-fixed frame
, respectively. 
Then the frictional work of the fluid and the molecule can be written as 
\begin{equation}
W=\zeta\int\d\hat{\bm{r}} \hat{\rho}(\hat{\bm{r}})|\bm{u}-\bm{u}_p|^2. \label{work_fric}
\end{equation}
Note that $\bm{u}_p=\bm{w}_{f}-\bm{g}\times\hat{\bm{r}}$. 
Since the scale of rigid molecule is much smaller than the fluid field, 
we may suppose that the flow is linear. 
In other words, if we denote $\kappa_{ij}=\partial_ju_i$, then we may assume 
$$
\bm{u}(\hat{\bm{r}})=\kappa\cdot\hat{\bm{r}}+\bm{u}_{0}. 
$$
%\kappa\hat{\bm{r}}_i=\kappa_{ij}\hat{\bm{r}}_j
Here, $\bm{u}_0$ is the velocity at $\bm{x}$, where $\hat{O}$ is located. 
By minimizing (\ref{work_fric}), we deduce that 
\begin{align}
\bm{w}_f=\bm{u}_0,\quad
\bm{g}=m_0\bm{I}^{-1}\int\d\hat{\bm{r}}\hat{\rho}(\hat{\bm{r}})\Big(\hat{\bm{r}}\times\kappa\cdot\hat{\bm{r}}\Big). \label{ang_v}
\end{align}
%Note that the point $\hat{O}$ is located at $\bm{x}$. 
If we denote the fluid velocity by $\bm{v}(\bm{x})$, then we may write 
\begin{equation}
  \bm{w}_f=\bm{u}_0=\bm{v}.    
\end{equation}

Summarizing the derivation above, the Smoluchowski equation can be rewritten as follows, 
\begin{align}
\frac{\partial f}{\partial t}+\nabla\cdot (f\bm{v})
= \nabla\cdot(\bm{J}(k_BT\nabla f+f\nabla V))
+L\cdot [(D_0\bm{I}^{-1})(k_BTLf+fLV)] - L\cdot(\bm{g}f).
\label{eqn_f}
\end{align}

\subsubsection{Momentum equation}
The incompressibility gives 
\begin{equation}
  \nabla\cdot\bm{v}=0. \label{incomp}
\end{equation}
The momentum equation is written as 
\begin{equation}
  \rho_s\left(\frac{\partial\bm{v}}{\partial t}+\bm{v}\cdot\nabla\bm{v}\right)
  =-\nabla p+\nabla\cdot\tau+\bm{F}_e, 
  \label{eqn_v}
\end{equation}
where $\rho_s$ is the density of the fluid, $\bm{F}_e$ is the external force, 
and $\tau=\tau_e+\tau_v$ is the stress, divided into the elastic and the 
viscous part. 
The elastic stress $\tau_e$ and the external force $\bm{F}_e$ can be derived 
from the principle of virtual work. 
Because the derivation is standard and can be found in literature \cite{Doi_book,cms2010}, 
we only list the results here. 
The external force $\bm{F}_e$ is given by 
\begin{equation}
  \bm{F}_e=-\int\d\nu~\nabla \mu(f)f=-c\left<\nabla \mu\right>, \label{extF}
\end{equation}
where we recall that $c$ is the concentration. For $\tau_e$, if we express $\bm{g}$ as 
\begin{equation}
  \bm{g}=\kappa_{jk}:(\alpha_i)_{jk}\bm{m}_i, 
\end{equation}
then 
\begin{equation}
  (\tau_e)_{jk}=c\left<(\alpha_i)_{jk}L_i\mu\right>. \label{tau_e}
\end{equation}
When $\bm{I}$ is diagonal, by (\ref{ang_v}), we deduce 
\begin{equation}
  \alpha_i=m_0I_{ii}^{-1}\int\d\hat{\bm{r}}\hat{\rho}(\hat{\bm{r}})\hat{\bm{r}}(\bm{m}_i\times\hat{\bm{r}}). 
\end{equation}
The viscous stress can be expressed as 
\begin{equation}
  \tau_v=2\eta(\kappa+\kappa^T)+\tau_{vf}. 
\end{equation}
The first term is the contribution of the friction in the fluid itself, 
and $\tau_{vf}$ is the contribution of the friction between the fluid and the molecules, determined by the following equation 
\begin{equation}
  c\left<W\right>=\kappa:\tau_{vf}, \label{str_fric}
\end{equation}
where $W$ is given by (\ref{work_fric}) with $\bm{g}$ taking (\ref{ang_v}). 

The whole system is described by (\ref{eqn_f})-(\ref{eqn_v}), with the terms 
given by (\ref{rot_diff}), (\ref{ang_v}), (\ref{extF}), (\ref{tau_e}), (\ref{str_fric}). 
It is worth noting that all the terms are derived from the distribution of sphere centers $\hat{\rho}(\hat{\bm{r}})$ and the free energy $F[f]$. 

\subsubsection{Energy dissipation law}
The energy of the system includes the 
free energy (\ref{FreeEng0}) and the kinetic energy of the fluid, 
$$
\int\d\bm{x}\, \frac{\rho_s}{2}|\bm{v}|^2+F[f]. 
$$
Now let us prove the energy dissipation law. 
For simplicity, we omit $k_BT$ in the following. We have 
\begin{align}
  &\frac{\d}{\d t}\left(\int\d\bm{x}\,\frac{\rho_s}{2}|\bm{v}|^2
  +F[f]\right)
  =\int\d\bm{x}\left(\rho_s\bm{v}\cdot\frac{\partial\bm{v}}{\partial t}
  +\int\d\nu\mu\frac{\partial f}{\partial t}\right)\nonumber\\
  =&\int\d\bm{x}\left[\int\d\nu\,
  \mu L\cdot(D_0\bm{I}^{-1}fL\mu)+\mu\nabla\cdot(\bm{J}f\nabla\mu)
  -\mu L\cdot(\bm{g}f)-\mu\nabla\cdot(\bm{v}f) \right]\nonumber\\
  &-\rho_s \bm{v}\cdot(\bm{v}\cdot\nabla)\bm{v}-\bm{v}\cdot\nabla p
  +\bm{v}\cdot(\nabla\cdot\tau)+\bm{v}\cdot\bm{F}_e\nonumber\\
  =&\int\d\bm{x}\bigg[\int\d\nu\,
  -f(L\mu)^TD_0\bm{I}^{-1}(L\mu)-f(\nabla\mu)^T\bm{J}(\nabla\mu)
  +f\bm{g}\cdot L\mu+\bm{v}\cdot f\nabla\mu\bigg]\nonumber\\
  &+\frac{\rho_s}{2} (\nabla\cdot\bm{v})|\bm{v}|^2+p\nabla\cdot\bm{v}
  -\kappa:(\tau_e+\tau_f)+\bm{v}\cdot\bm{F}_e
  \nonumber\\
  =&\int\d\bm{x}\bigg[\int\d\nu\,
  -f(L\mu)^TD_0\bm{I}^{-1}(L\mu)-f(\nabla\mu)^T\bm{J}(\nabla\mu)\nonumber\\
  &+(f\bm{g}\cdot L\mu-f(\kappa:\alpha_j)L_j\mu)
  +(\bm{v}\cdot f\nabla\mu-\bm{v}\cdot c\rho\nabla\mu)
  \bigg]-\kappa:\tau_f\nonumber\\
  =&\int\d\bm{x}\left[
  -\left<(L\mu)^TD_0\bm{I}^{-1}(L\mu)\right>
  -\left<(\nabla\mu)^T\bm{J}(\nabla\mu)\right>
  -2\eta\frac{\kappa+\kappa^T}{2}:\frac{\kappa+\kappa^T}{2}
  -\kappa:\tau_{vf}\right]. 
\end{align}
In the above, we ignore the boundary terms. 
Note that the first three terms are not positive. By (\ref{str_fric}) and 
(\ref{work_fric}) we know that the last term is not positive either. 

\subsection{Bent-core molecules and star molecules}
For bent-core molecules and star molecules, the sphere centers are distributed in the plane $\hat{O}\bm{m}_1\bm{m}_2$. In this case, we can simplify the expressions derived above. 
First, we have $I_{33}=I_{11}+I_{22}$. 
Thus, in (\ref{ang_v}), 
\begin{equation}
  \bm{g}=(\kappa:\bm{m}_2\bm{m}_3)\bm{m}_1-(\kappa:\bm{m}_1\bm{m}_3)\bm{m}_2
  +\frac{1}{I_{11}+I_{22}}
  (I_{22}\kappa:\bm{m}_1\bm{m}_2
  -I_{11}\kappa:\bm{m}_2\bm{m}_1)\bm{m}_3. \label{grot_bent}
\end{equation}
Then in (\ref{tau_e}), we have 
\begin{align}
  \alpha_1&=\bm{m}_2\bm{m}_3,\ \alpha_2=-\bm{m}_1\bm{m}_3,\ 
  \alpha_3=\frac{1}{I_{11}+I_{22}}
  (I_{22}\bm{m}_1\bm{m}_2 - I_{11}\bm{m}_2\bm{m}_1).
\end{align}
By (\ref{work_fric}) and (\ref{str_fric}), we deduce that 
\begin{align}
  \tau_{vf}=&\frac{c\zeta\kappa}{m_0}:\Big[I_{22}\left<\bm{m}_1\bm{m}_1\bm{m}_1\bm{m}_1\right>
    +I_{11}\left<\bm{m}_2\bm{m}_2\bm{m}_2\bm{m}_2\right>\nonumber\\
    &+\frac{I_{11}I_{22}}{I_{11}+I_{22}}\left<(\bm{m}_1\bm{m}_2+\bm{m}_2\bm{m}_1)
    (\bm{m}_1\bm{m}_2+\bm{m}_2\bm{m}_1)\right>\Big]. \label{tau_vf}
\end{align}

\begin{figure}
  \centering
  \includegraphics[width=0.3\textwidth,keepaspectratio]{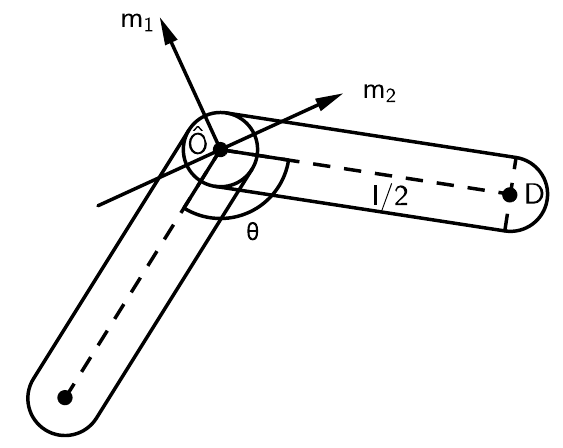}
  \includegraphics[width=0.3\textwidth,keepaspectratio]{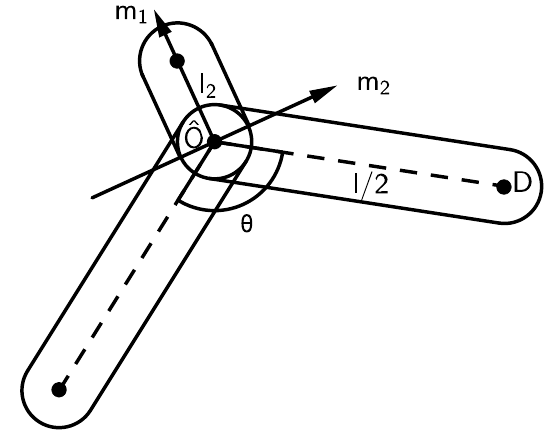}
  \caption{Bent-core molecule and star molecule. }
  \label{mol}
\end{figure}
We can see that the only difference in the above terms lies in the coefficients as functions of the moment of inertia, from which we can distinguish the bent-core molecules and star molecules. 
For a bent-core molecule (drawn in Fig. \ref{mol} left), 
the sphere centers are distributed uniformly and continuously on a 
two-segment broken line, where the length of each segment is $l/2$. 
Thus, $\hat{\rho}$ is given by 
\begin{equation}
\hat{\rho}(\bm{\hat{r}})=\frac{1}{l}\int_{-\frac{l}{2}}^{\frac{l}{2}}\d s~
\delta\Big(\bm{\hat{r}}-(\frac{l}{4}-|s|)\cos\frac{\theta}{2}\bm{m}_1
  -s\sin\frac{\theta}{2}\bm{m}_2\Big). \label{sphdis}
\end{equation}
Substituting it into (\ref{MoI}) and recalling (\ref{rot_diff}), we obtain
\begin{equation}
  I_{11}=\frac{l^2m_0}{48}\cdot 4\sin^2\frac{\theta}{2},\quad
  I_{22}=\frac{l^2m_0}{48}\cdot \cos^2\frac{\theta}{2}.
\end{equation}
For a star molecule (drawn in Fig. \ref{mol} right), the sphere centers also lie in a third line segment of the length $l_2$. 
Thus, $\hat{\rho}$ is given by 
\begin{equation}
\hat{\rho}(\bm{\hat{r}})=\frac{1}{l+l_2}\left[\int_{-\frac{l}{2}}^{\frac{l}{2}}\d s\,
\delta\Big(\bm{\hat{r}}-(\frac{l}{4}-|s|)\cos\frac{\theta}{2}\bm{m}_1
  -s\sin\frac{\theta}{2}\bm{m}_2\Big)
+\int_{0}^{l_2}\d s\,
\delta\Big(\bm{\hat{r}}-s\bm{m}_1\Big)\right]. \label{sphdis_star}
\end{equation}
Therefore, 
\begin{align}
  I_{11}=\frac{l^2m_0}{12}\sin^2\frac{\theta}{2},\quad
  I_{22}=m_0\left(\frac{\frac{1}{3}l_2^3+\frac{1}{12}l^3\cos^2\frac{\theta}{2}}{l+l_2}-x_C^2\right), 
\end{align}
where
$$
x_C=\frac{\frac{1}{2}l_2^2-\frac{1}{4}l^2\cos\frac{\theta}{2}}{l+l_2} 
$$
is the $\bm{m}_1$-coordinate of the center of mass. 

\begin{figure}
  \centering
  \includegraphics[width=0.4\textwidth,keepaspectratio]{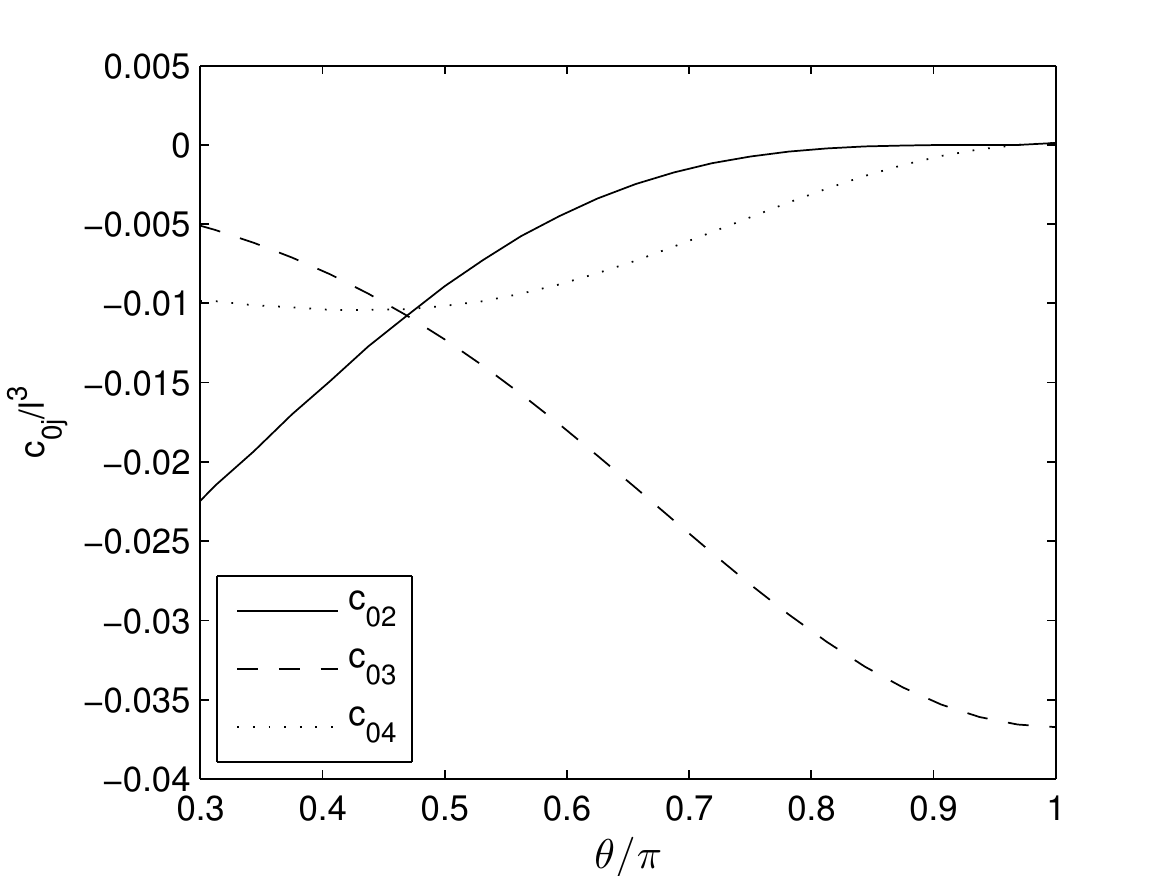}
  \includegraphics[width=0.4\textwidth,keepaspectratio]{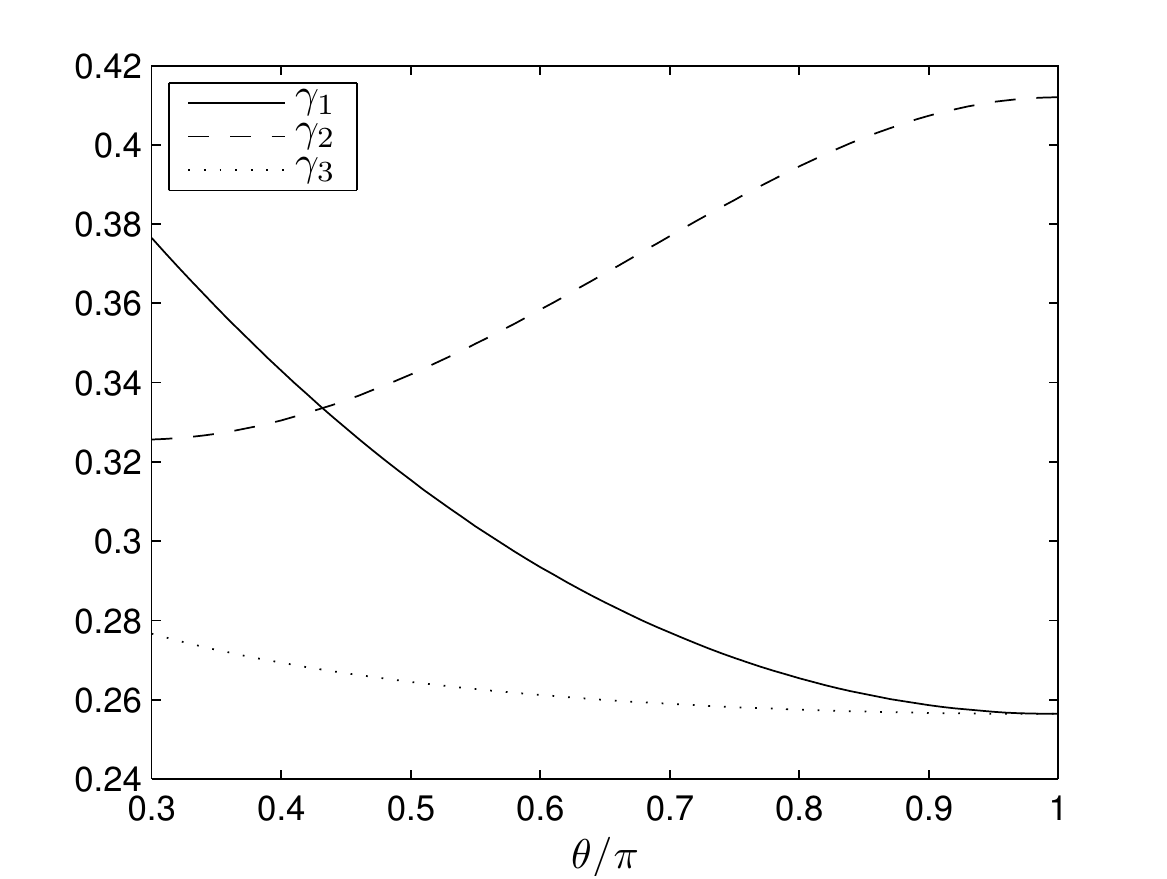}
  \caption{Some coefficients in the model for bent-core molecules with $D/l=1/40$. Left: coefficients of the bulk energy. Right: translational diffusion coefficients. }
  \label{difcoe}
\end{figure}

For bent-core molecules and star molecules, the spatial diffusion matrix $\bm{J}$ derived from the Kirkwood theory (see Appendix) is diagonal in the frame $(\bm{m}_i)$, 
\begin{equation}
  \bm{J}=\frac{1}{8\pi D\eta_0}\sum_{j=1}^3\gamma_j\bm{m}_j\bm{m}_j. \label{Jcoef}
\end{equation}
For bent-core molecules with $D/l=1/40$, we plot $\gamma$ in Fig. \ref{difcoe} right. 

For the free energy, we adopt the tensor model derived in \cite{BentModel} from the second virial expansion with the hardcore molecular interaction. 
The hardcore interaction is determined only by the molecular shape, given by $\hat{\rho}$ in the current context. 
Assume that the concentration $c$ is constant. 
Define $\bm{p}=\left<\bm{m}_1\right>$, $Q_1=\left<\bm{m}_1\bm{m}_1\right>$,
$Q_2=\left<\bm{m}_2\bm{m}_2\right>$.
For both molecules, $F_{r}$ shares the form below, 
\begin{align}
  \frac{F_r}{k_BT}
  =&\int\d\bm{x}\,\frac{c^2}{2}(c_{01}|\bm{p}|^2+c_{02}|Q_1|^2+c_{03}|Q_2|^2
  +2c_{04}Q_1:Q_2)\nonumber\\
  &+c^2(c_{11}p_j\partial_iQ_{1ij}
  +c_{12}p_j\partial_iQ_{2ij})\nonumber\\
  &+\frac{c^2}{4}\left[c_{21}|\nabla \bm{p}|^2
  + c_{22} |\nabla Q_1|^2 + c_{23}|\nabla Q_2|^2
  + 2c_{24}\partial_{i}Q_{1jk}\partial_iQ_{2jk}\right.\nonumber\\
  &+ 2c_{27}\partial_{i}p_i\partial_jp_j
  + 2c_{28}\partial_iQ_{1ik}\partial_jQ_{1jk}
  + \left.2c_{29}\partial_iQ_{2ik}\partial_jQ_{2jk}
  + 4c_{2,10}\partial_iQ_{1ik}\partial_jQ_{2jk}\right], \label{FreeEng}
\end{align}
which is determined by the molecular symmetry \cite{BentModel,SymmO}. 
The difference also lies in the coefficients $c_{kj}$. 
They are derived as functions of the molecular parameters $l$, $D$, 
$\theta$ for bent-core molecules, and also $l_2$ for star molecules. 
They possess the scaling property 
\begin{equation}
  c_{kj}=l^{k+3}\tilde{c}_{kj}(D/l,\theta,l_2/l). \label{scaling} 
\end{equation}
The calculation of $c_{kj}$ is discussed in \cite{BentModel,SymmO}. 
In Fig. \ref{difcoe} left we plot $c_{02},c_{03},c_{04}$ that are necessary for the shear flow problem. 

To summarize, we establish the dynamic model from the molecular shape described by $\hat{\rho}$, and the free energy $F[f]$. In the case of hardcore interaction, $F[f]$ is also determined by the molecular shape. Therefore, the model is able to characterize the dynamics of molecules with different shapes. 
In particular, for bent-core molecules and star molecules, the model has the same form, no matter for the free energy $F$ and other terms. 
The two types of molecules are distinguished by numerous coefficients in the model, which are expressed as functions of molecular parameters. 

\section{Tensor model\label{tensormodel}}
We derive the tensor model from the molecular model. 
When we use the free energy \eqref{FreeEng}, the elastic stress $\tau_e$ and the external force $\bm{F}_e$ can also be expressed by tensors. 
Let $V$ be computed from the free energy (\ref{FreeEng}). 
Denote 
$$
V_{\bm{p}}=\frac{1}{k_BT}\cdot\frac{\delta F_r}{\delta \bm{p}},\ V_{Q_1}=\frac{1}{k_BT}\cdot\frac{\delta F_r}{\delta Q_1},\ V_{Q_2}=\frac{1}{k_BT}\cdot\frac{\delta F_r}{\delta Q_2}. 
$$
Direct computation gives 
\begin{align}
  V_{\bm{p}}=&c_{01}\bm{p}+c_{11}\nabla\cdot Q_1+c_{12}\nabla\cdot Q_2-c_{21}\Delta\bm{p}-c_{27}\nabla(\nabla\cdot\bm{p}),\label{V_begin}\\
  V_{Q_1}=&c_{02}Q_1+c_{04}Q_2-c_{11}\nabla\bm{p}-c_{22}\Delta Q_1-c_{24}\Delta Q_2-c_{28}\nabla(\nabla\cdot Q_1)-c_{2,10}\nabla(\nabla\cdot Q_2),\\
  V_{Q_2}=&c_{04}Q_1+c_{03}Q_2-c_{12}\nabla\bm{p}-c_{24}\Delta Q_1-c_{23}\Delta Q_2-c_{2,10}\nabla(\nabla\cdot Q_1)-c_{29}\nabla(\nabla\cdot Q_2). \label{V_end}
\end{align}
And we can verify that 
\begin{equation}
  V=V_{\bm{p}}\cdot\bm{m}_1+V_{Q_1}:\bm{m}_1\bm{m}_1+V_{Q_2}:\bm{m}_2\bm{m}_2. \label{Vtensor}
\end{equation}
Thus 
\begin{equation}
  L_iV=V_{\bm{p}}\cdot(L_i\bm{m}_1)+V_{Q_1}:(L_i\bm{m}_1\bm{m}_1)
  +V_{Q_2}:(L_i\bm{m}_2\bm{m}_2).
\end{equation}
From this equation, the elastic stress can be written as 
\begin{align}
  \tau_e=&ck_BT\big\{
  \left<\bm{m}_2\bm{m}_2-\bm{m}_3\bm{m}_3\right>
  +V_{Q_2}:\left<(\bm{m}_2\bm{m}_3+\bm{m}_3\bm{m}_2)\bm{m}_2\bm{m}_3\right>\nonumber\\
  &+\left<\bm{m}_1\bm{m}_1-\bm{m}_3\bm{m}_3\right>
  +V_{\bm{p}}\cdot\left<\bm{m}_3\bm{m}_1\bm{m}_3\right>
  +V_{Q_1}:\left<(\bm{m}_1\bm{m}_3+\bm{m}_3\bm{m}_1)\bm{m}_1\bm{m}_3\right>\nonumber\\
  &+\frac{1}{I_{11}+I_{22}}\big[
    (I_{22}-I_{11})\left<\bm{m}_2\bm{m}_2-\bm{m}_1\bm{m}_1\right>
    +V_{\bm{p}}\cdot\left<\bm{m}_2(I_{22}\bm{m}_1\bm{m}_2-I_{11}\bm{m}_2\bm{m}_1)
    \right>\nonumber\\
  &
    +(V_{Q_1}-V_{Q_2}):\left<(\bm{m}_1\bm{m}_2+\bm{m}_2\bm{m}_1)
    (I_{22}\bm{m}_1\bm{m}_2-I_{11}\bm{m}_2\bm{m}_1)\right>\big]\big\}. \label{tau_et}
\end{align}
And the external force is written as 
\begin{equation}
  \bm{F}_e=-ck_BT\nabla(V_{\bm{p}}\cdot\bm{p}+V_{Q_1}:Q_1+V_{Q_2}:Q_2). \label{extFt}
\end{equation}
Now the equation (\ref{eqn_v}) is only relevant to the tensors. 
For the equation (\ref{eqn_f}), we multiply it by the tensors and integrate in 
$SO_3$. Generally, we can write 
\begin{align}
\frac{\partial A}{\partial t}+\bm{v}\cdot\nabla A
= \mathcal{N}_{A} + \mathcal{M}_{A} + \mathcal{V}_{A}, \label{tens}
\end{align}
where $A$ is arbitrary tensor, and $\mathcal{N}_{A},~\mathcal{M}_{A},~\mathcal{V}_{A}$
are the terms computed from spatial diffusion terms, 
rotational diffusion terms, and rotational convection terms. 
We need the integration by parts (\ref{IntPart}) and (\ref{diffL}) when computing these terms. 
Take $Q_1$ as an example. After multiplying $\bm{m}_1\bm{m}_1$ and doing the integration, the following term appears, 
\begin{align}
\int\d\nu \bm{m}_1\bm{m}_1D_2L_2^2f=&
D_2\int\d\nu L_2^2(\bm{m}_1\bm{m}_1)f\nonumber\\
=&D_2\int\d\nu 2(\bm{m}_3\bm{m}_3-\bm{m}_1\bm{m}_1)f
=2D_2\left<\bm{m}_3\bm{m}_3-\bm{m}_1\bm{m}_1\right>. \label{ComputeRight}
\end{align}
Similarly, we can derive that 
\begin{align}
  -\mathcal{M}_{\bm{p}}=&D_2\left[\bm{p}
  +V_{\bm{p}}\cdot\left<\bm{m}_3\bm{m}_3\right>
  +V_{Q_1}:(\left<\bm{m}_1\bm{m}_3\bm{m}_3\right>
  +\left<\bm{m}_3\bm{m}_1\bm{m}_3\right>)\right]\nonumber\\
  &+D_3\left[\bm{p}
  +V_{\bm{p}}\cdot\left<\bm{m}_2\bm{m}_2\right>
  +(V_{Q_1}-V_{Q_2}):(\left<\bm{m}_1\bm{m}_2\bm{m}_2\right>
  +\left<\bm{m}_2\bm{m}_1\bm{m}_2\right>)\right].
  \label{tensor_begin}\\
  -\mathcal{M}_{Q_1}=&D_2\left[2(Q_1-Q_3)
  +V_{\bm{p}}\cdot(\left<\bm{m}_3\bm{m}_3\bm{m}_1\right>
  +\left<\bm{m}_3\bm{m}_1\bm{m}_3\right>)\right.\nonumber\\
  &+\left.V_{Q_1}:\left<(\bm{m}_1\bm{m}_3+\bm{m}_3\bm{m}_1)
  (\bm{m}_1\bm{m}_3+\bm{m}_3\bm{m}_1)\right>\right]\nonumber\\
  &+D_3\left[2(Q_1-Q_2)
  +V_{\bm{p}}\cdot(\left<\bm{m}_2\bm{m}_2\bm{m}_1\right>
  +\left<\bm{m}_2\bm{m}_1\bm{m}_2\right>)\right.\nonumber\\
  &\left.+(V_{Q_1}-V_{Q_2}):\left<(\bm{m}_1\bm{m}_2+\bm{m}_2\bm{m}_1)
  (\bm{m}_1\bm{m}_2+\bm{m}_2\bm{m}_1)\right>\right].\\
%  \\
  -\mathcal{M}_{Q_2}=&D_1[2(Q_2-Q_3)
  +V_{Q_2}:\left<(\bm{m}_2\bm{m}_3+\bm{m}_3\bm{m}_2)
  (\bm{m}_2\bm{m}_3+\bm{m}_3\bm{m}_2)\right>]
  \nonumber\\
  &+D_3\left[2(Q_2-Q_1)
  -V_{\bm{p}}\cdot(\left<\bm{m}_2\bm{m}_2\bm{m}_1\right>
  +\left<\bm{m}_2\bm{m}_1\bm{m}_2\right>)\right.\nonumber\\
  &\left.-(V_{Q_1}-V_{Q_2}):\left<(\bm{m}_1\bm{m}_2+\bm{m}_2\bm{m}_1)
  (\bm{m}_1\bm{m}_2+\bm{m}_2\bm{m}_1)\right>\right].\\
%  \\
  \mathcal{V}_{\bm{p}}=&\kappa:\big[\left<\bm{m}_1\bm{m}_3\bm{m}_3\right>
  +\frac{I_{22}}{I_{11}+I_{22}}\left<\bm{m}_1\bm{m}_2\bm{m}_2\right>
  -\frac{I_{11}}{I_{11}+I_{22}}\left<\bm{m}_2\bm{m}_1\bm{m}_2\right>\big].\\
%  \\
  \mathcal{V}_{Q_1}=&\kappa:\big[\left<\bm{m}_1\bm{m}_3(\bm{m}_1\bm{m}_3+\bm{m}_3\bm{m}_1)\right>%\nonumber\\
  +\frac{I_{22}}{I_{11}+I_{22}}\left<\bm{m}_1\bm{m}_2(\bm{m}_1\bm{m}_2+\bm{m}_2\bm{m}_1)\right>\nonumber\\
  &-\frac{I_{11}}{I_{11}+I_{22}}\left<\bm{m}_2\bm{m}_1(\bm{m}_1\bm{m}_2+\bm{m}_2\bm{m}_1)\right>\big].\\
%  \\
  \mathcal{V}_{Q_2}=&\kappa:\big[\left<\bm{m}_2\bm{m}_3(\bm{m}_2\bm{m}_3+\bm{m}_3\bm{m}_2)\right>%\nonumber\\
  -\frac{I_{22}}{I_{11}+I_{22}}\left<\bm{m}_1\bm{m}_2(\bm{m}_1\bm{m}_2+\bm{m}_2\bm{m}_1)\right>\nonumber\\
  &+\frac{I_{11}}{I_{11}+I_{22}}\left<\bm{m}_2\bm{m}_1(\bm{m}_1\bm{m}_2+\bm{m}_2\bm{m}_1)\right>\big].\\
  \left(\mathcal{N}_{\bm{p}}\right)_{\alpha}=&
  \partial_i\left(\partial_j\sum_{\sigma=1}^3\gamma_{\sigma}
  \left<m_{1\alpha}m_{\sigma i}m_{\sigma j}\right>\right.%\nonumber\\
  +\partial_j(V_{\bm{p}})_k\sum_{\sigma=1}^3\gamma_{\sigma}
  \left<m_{1k}m_{1\alpha}m_{\sigma i}m_{\sigma j}\right>\nonumber\\
  &+\partial_j(V_{Q_1})_{kl}\sum_{\sigma=1}^3\gamma_{\sigma}
  \left<m_{1k}m_{1l}m_{1\alpha}m_{\sigma i}m_{\sigma j}\right>
  \nonumber\\
  &\left.+\partial_j(V_{Q_2})_{kl}\sum_{\sigma=1}^3\gamma_{\sigma}
  \left<m_{2k}m_{2l}m_{1\alpha}m_{\sigma i}m_{\sigma j}\right>
  \right).\\
  \left(\mathcal{N}_{Q_1}\right)_{\alpha\beta}=&
  \partial_i\left(\partial_j\sum_{\sigma=1}^3\gamma_{\sigma}
  \left<m_{1\alpha}m_{1\beta}m_{\sigma i}m_{\sigma j}\right>\right.%\nonumber\\
  +\partial_j(V_{\bm{p}})_k\sum_{\sigma=1}^3\gamma_{\sigma}
  \left<m_{1k}m_{1\alpha}m_{1\beta}m_{\sigma i}m_{\sigma j}\right>\nonumber\\
  &+\partial_j(V_{Q_1})_{kl}\sum_{\sigma=1}^3\gamma_{\sigma}
  \left<m_{1k}m_{1l}m_{1\alpha}m_{1\beta}m_{\sigma i}m_{\sigma j}\right>\nonumber\\
  &\left.+\partial_j(V_{Q_2})_{kl}\sum_{\sigma=1}^3\gamma_{\sigma}
  \left<m_{2k}m_{2l}m_{1\alpha}m_{1\beta}m_{\sigma i}m_{\sigma j}\right>\right).\\
  \left(\mathcal{N}_{Q_2}\right)_{\alpha\beta}=&
  \partial_i\left(\partial_j\sum_{\sigma=1}^3\gamma_{\sigma}
  \left<m_{2\alpha}m_{2\beta}m_{\sigma i}m_{\sigma j}\right>\right.%\nonumber\\
  +\partial_j(V_{\bm{p}})_k\sum_{\sigma=1}^3\gamma_{\sigma}
  \left<m_{1k}m_{2\alpha}m_{2\beta}m_{\sigma i}m_{\sigma j}\right>\nonumber\\
  &+\partial_j(V_{Q_1})_{kl}\sum_{\sigma=1}^3\gamma_{\sigma}
  \left<m_{1k}m_{1l}m_{2\alpha}m_{2\beta}m_{\sigma i}m_{\sigma j}\right>\nonumber\\
  &\left.+\partial_j(V_{Q_2})_{kl}\sum_{\sigma=1}^3\gamma_{\sigma}
  \left<m_{2k}m_{2l}m_{2\alpha}m_{2\beta}m_{\sigma i}m_{\sigma j}\right>\right).  \label{tensor_end}
\end{align}

\iffalse
Denote the high-order tensors that appear in the above by 
\begin{align*}
  T_2=&\left<\bm{m}_1\bm{m}_2\bm{m}_2\right>,\\
  T_3=&\left<\bm{m}_1\bm{m}_3\bm{m}_3\right>,\\
  S_1=&\left<\bm{m}_1\bm{m}_1\bm{m}_1\bm{m}_1\right>,\\
  S_2=&\left<\bm{m}_2\bm{m}_2\bm{m}_2\bm{m}_2\right>,\\
  R_2=&\left<\bm{m}_1\bm{m}_1\bm{m}_2\bm{m}_2\right>,\\
  R_3=&\left<\bm{m}_1\bm{m}_1\bm{m}_3\bm{m}_3\right>,\\
  R_4=&\left<\bm{m}_2\bm{m}_2\bm{m}_3\bm{m}_3\right>.
\end{align*}
It holds the following relations for these tensors, 
\begin{align}
  R_3=&Q_1I-S_1-R_2,\label{HTensor1}\\
  R_{4ijkl}=&Q_{2ij}\delta_{kl}-S_{2ijkl}-R_{2klij}. \label{HTensor2}
\end{align}
\fi

To make the equations form a closed system, we need to express high-order tensors as functions of $(\bm{p},Q_1,Q_2)$. 
Here we use the quasi-equilibrium approximation, namely to choose $f$ that minimizes the entropy term $\int \d\nu f\log f$ 
with $(\bm{p},Q_1,Q_2)$ equal to the given value. 
Remember that $f=c\rho$ where $c$ is constant. Thus $\rho$ is given by \cite{BentModel}
\begin{equation}
  \rho(P)=\frac{1}{Z}\exp (\bm{b}\cdot\bm{m}_1 
  + B_1:\bm{m}_1\bm{m}_1 + B_2:\bm{m}_2\bm{m}_2), \label{form1}
\end{equation}
where $\bm{b}$ is a vector, $B_1$ and $B_2$ are symmetric matrices, and 
\begin{equation}
Z=\int\d\nu \exp (\bm{b}\cdot\bm{m}_1
  + B_1:\bm{m}_1\bm{m}_1 + B_2:\bm{m}_2\bm{m}_2). \label{PartFunc}
\end{equation}

Now the system is described by $(\bm{p},Q_1,Q_2)$. The evolution of three tensors is governed by (\ref{tens}) in which the terms are given by (\ref{tensor_begin})-(\ref{tensor_end}), 
together with (\ref{incomp}) and (\ref{eqn_v}) in which the terms are given by (\ref{tau_vf}), (\ref{tau_et}), (\ref{extFt}). 
The high-order tensors are computed from (\ref{form1}), which keeps $f$ positive 
and the energy dissipation law. 
It is obvious that $f$ is positive, and we can deduce that 
\begin{align}
  &\frac{\d}{\d t}\left(F[\bm{p},Q_1,Q_2]+
    \int\d\bm{x}\frac{\rho|\bm{v}|^2}{2}\right)\nonumber\\
 =&\int\d\bm{x}\left\{
 -ck_BT\left[D_1\left<(\mu_{Q_2}:(\bm{m}_2\bm{m}_3+\bm{m}_3\bm{m}_2))^2\right>\right.\right.\nonumber\\
   &+D_2\left<(\mu_{\bm{p}}\cdot\bm{m}_3
   +\mu_{Q_1}:(\bm{m}_2\bm{m}_3+\bm{m}_3\bm{m}_2))^2\right>\nonumber\\
   &\left.+D_3\left<(\mu_{\bm{p}}\cdot\bm{m}_2
   +(\mu_{Q_1}-\mu_{Q_2}):(\bm{m}_2\bm{m}_3+\bm{m}_3\bm{m}_2))^2\right>
   \right]\nonumber\\
 &-ck_BT\sum_{\sigma=1}^3\gamma_{\sigma}
  \left<[\partial_j(\mu_{\bm{p}})_km_{1k}m_{\sigma j}+\partial_j(\mu_{Q_1})_{kl}m_{1k}m_{1l}m_{\sigma j}+\partial_j(\mu_{Q_2})_{kl}m_{2k}m_{2l}m_{\sigma j}]^2\right>
\nonumber\\
 &-2\eta\frac{\kappa+\kappa^T}{2}:\frac{\kappa+\kappa^T}{2}%\nonumber\\
 -c\zeta\bigg[I_{22}\left<(\kappa:\bm{m}_1\bm{m}_1)^2\right>
   +I_{11}\left<(\kappa:\bm{m}_2\bm{m}_2)^2\right>\nonumber\\
 &\left.+\frac{I_{11}I_{22}}{I_{11}+I_{22}}\left<(\kappa:(\bm{m}_1\bm{m}_2+\bm{m}_2\bm{m}_1))^2\right>
   \bigg]\right\}. \label{disp}
\end{align}
In the above, we denote $\mu_{\bm{p}}=\delta F/\delta \bm{p}$, etc. 
The details are given in Appendix. 
Note that each of the right-hand terms is not positive. 

\section{Numerical results\label{results}}
In this section, we focus on the shear flow problem. 
We assume that the velocity is along the $x$-direction, 
and the gradient is along the $y$-direction, and 
$$
\kappa_{12}=\partial_2v_1=k 
$$
is a constant. 
We also assume that the tensors are spatially homogeneous. 
Thus, we only need the bulk energy in \eqref{FreeEng} and will discard the gradient terms. 
In this case, the equation of momentum holds naturally, and we need to solve the Smoluchowski equation only. 

We rescale the time unit by $\tilde{t}=(\zeta l^2/48k_BT)^{-1}t$. 
It cancels the $k_BT$ in the free energy and the units in the rotational diffusion coefficients $D_i$. 
For bent-core molecules, the rescaling let 
$$
(D_1^{-1},D_2^{-1},D_3^{-1})=(4\sin^2\frac{\theta}{2}, \cos^2\frac{\theta}{2}, 1+3\sin^2\frac{\theta}{2}).
$$
After the rescaling, the shear rate $k$ becomes dimensionless. 
In the free energy, by (\ref{scaling}) we rescale $\tilde{\bm{x}}=\bm{x}/l$ 
and reduce the shape parameters to three dimensionless ones: 
$\eta=D/l$, $l_2/l$, and $\theta$. 
We fix $\eta=1/40$, and express the concentration by $\alpha=cD^2(l+l_2)$ that is proportional to the volume fraction $(\pi/4) cD^2(l+l_2)$. 

In what follows, we examine both the molecular model and the tensor model, and compare the flowing modes shown by both models. 

\subsection{Numerical method}
For the tensor model, we use (\ref{form1}) to convert the equations of tensors into equations of 
$(\bm{b},B_1,B_2)$, 
\begin{align}
  \frac{\d (\bm{b},B_1,B_2)}{\d t}=
  \left(\frac{\partial (\bm{p},Q_1,Q_2)}{\partial (\bm{b},B_1,B_2)}\right)^{-1}
  \frac{\d (\bm{p},Q_1,Q_2)}{\d t}.
\end{align}
The derivatives $\frac{\partial (\bm{p},Q_1,Q_2)}{\partial (\bm{b},B_1,B_2)}$  
are computed by (\ref{covariance}). They are expressed by high-order tensors. 
The tensors are computed by numerical integration about the Euler angles. 
Each of the Euler angles are discretized by $32$ points. 
The time discretization is implemented by the classical fourth-order Runge-Kutta method with the time step $\delta t=10^{-2}$. 
The initial value is chosen as $B_1=B_2=0$, while $\bm{b}=(1.4,2.8,1.4)^T$ pointing to a tilted direction. 

For the molecular model, we adopt a spectral-Galerkin method, where we use Wigner D-matrix $D^j_{mm'}$ (see, for example, \cite{Wigner_book}), truncated at $j\le 10$, to discretize the density function $f$. 
%We refer to \cite{} for the definition and properties of Wigner D-matrix. 
For the time integration, we also use the classical fourth-order Runge-Kutta method, with the time step $\delta t=5\times 10^{-3}$. 
For the initial value, we start from the Boltzmann distribution with $B_1=B_2=0$ and $\bm{b}=(1.4,2.8,1.4)^T$. 
We let it evolve $2000$ time steps under the parameter $\alpha=0.5$, $\theta=23\pi/32$, $k=6.4$, and take the result as the initial value. 

\subsection{Flow modes}
%examine $\alpha=16.8,\ 20.0$, and let $k$, $\theta$ vary. 
Before looking at the flow modes, we review the homogeneous nematic phases shown by bent-core molecules and star molecules in quiescent fluid, namely $k=0$, which are discussed in \cite{BentModel}. 
In these phases we have $\bm{p}=0$. Denote $Q_3=\left<\bm{m}_3\bm{m}_3\right>=I-Q_1-Q_2$. 
Bent-core molecules and star molecules can exhibit the uniaxial nematic phase, where we can find a unit vector $\bm{n}$ such that 
$$
Q_i=s_i(\bm{n}\bm{n}-\frac{I}{3})+\frac{I}{3},\quad i=1,2,3. 
$$
According to the signs of $s_i$, the uniaxial nematic phase is further classified. 
It is the $N_2$ phase where $s_1,s_3<0$, $s_2>0$, indicating that $\bm{m}_2$ accumulates near $\bm{n}$ and $\bm{m}_1,\bm{m}_3$ accumulate near the plane vertical to $\bm{n}$; 
and the $N_3$ phase where $s_1,s_2<0$, $s_3>0$, indicating that $\bm{m}_3$ accumulates near $\bm{n}$ and $\bm{m}_2,\bm{m}_3$ accumulate near the plane vertical to $\bm{n}$. 
We can also observe the biaxial nematic phase ($B$), where we can find an orthonormal frame $(\bm{n}_1,\bm{n}_2,\bm{n}_3)$ such that 
$$
Q_i=s_{i1}\bm{n}_1\bm{n}_1+s_{i2}\bm{n}_2\bm{n}_2+s_{i3}\bm{n}_3\bm{n}_3. 
$$
The eigenvalues satisfy $s_{ii}>s_{ij}\, (j\ne i)$, indicating that $\bm{m}_i$ is preferably along $\bm{n}_i$. 

Both bent-core molecules and star molecules exhibit the isotropic phase with small $\alpha$, and the modulated twist-bend phase with large $\alpha$. 
Thus, in this work we will choose intermediate $\alpha$ to let the system have homogeneous nematic phases in equilibrium. 
%In particular, we choose $\alpha=0.42$ and $\alpha=0.5$, 
For bent-core molecules, we choose $\alpha=0.42,\, 0.5$, and examine the bending angles $\theta=j\pi/32$ where $16\le j\le 23$. 
For both $\alpha$, it shows the $N_2$ phase for $20\le j\le 23$, 
the $N_3$ phase for $16\le j\le 18$, and the $B$ phase for $j=19$. 
For star molecules, we fix $\alpha=0.42$, $\theta=2\pi/3$ and examine $l_2/l=j/40$ where $5\le j\le 11$. 
It shows the $N_2$ phase when $j=5$, the $N_3$ phase when $j=11$, and the $B$ phase when $6\le j\le 10$. 
%{\color{red}(PHASE DIAGRAM)}

We choose the shear rates as $k=0.2n,\, n=1,\ldots,100$. 
Since $\bm{p}$ is zero in quiescent fluid, we are interested in whether $\bm{p}$ appears. 
Actually, under our choice of parameters, $|\bm{p}|$ always decays rapidly. 
This should be the result of $c_{01}>0$ (see \cite{SymmO}). 
In the following, we no longer look at $\bm{p}$ and focus on $Q_1$ and $Q_2$. 
\iffalse
\begin{figure}
  \centering
  \includegraphics[width=0.5\textwidth,keepaspectratio]{pzero.pdf}
  \caption{$|\bm{p}|^2$ as function of $t$. }
  \label{pzero}
\end{figure}
\fi
Denote the unit eigenvector of the largest eigenvalue of $Q_i$ as $\bm{q}_i$ for $i=1,2$. 
Note that in quiescent fluid $\bm{q}_1$ and $\bm{q}_2$ are vertical. 
Although it does not hold in shear flow, we find that $\bm{q}_1$ and $\bm{q}_2$ are always approximately vertical. 
Actually, in most cases, we have $\cos\langle\bm{q}_1,\bm{q}_2\rangle\le 0.1$. 
This value may become a little larger when the shear rate $k$ is near the transition value between two flow modes. At that time, the largest and second eigenvalues of $Q_1$ or $Q_2$ might be very close so that $\bm{q}_1$ and $\bm{q}_2$ are sensitive to the value of $Q_1$ and $Q_2$. 
Thus, we may view the molecule as having a preferred orientation such that $\bm{m}_i$ is near $\bm{q}_i$, and classify the flow modes according to the motion of $\bm{q}_i$. 

\subsubsection{Molecular model}
\begin{figure}
  \centering
  \includegraphics[width=0.48\textwidth,keepaspectratio]{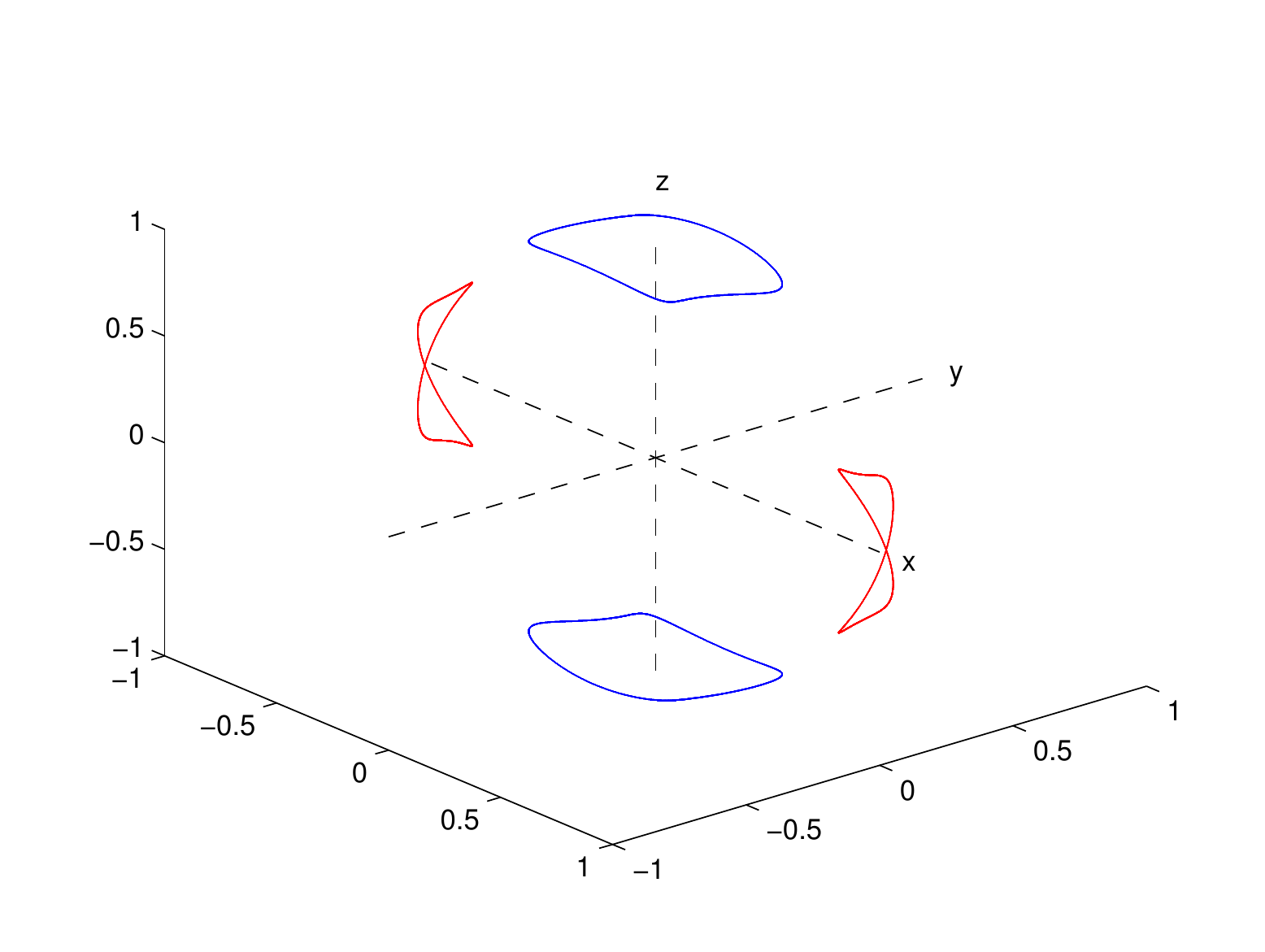}
  \includegraphics[width=0.48\textwidth,keepaspectratio]{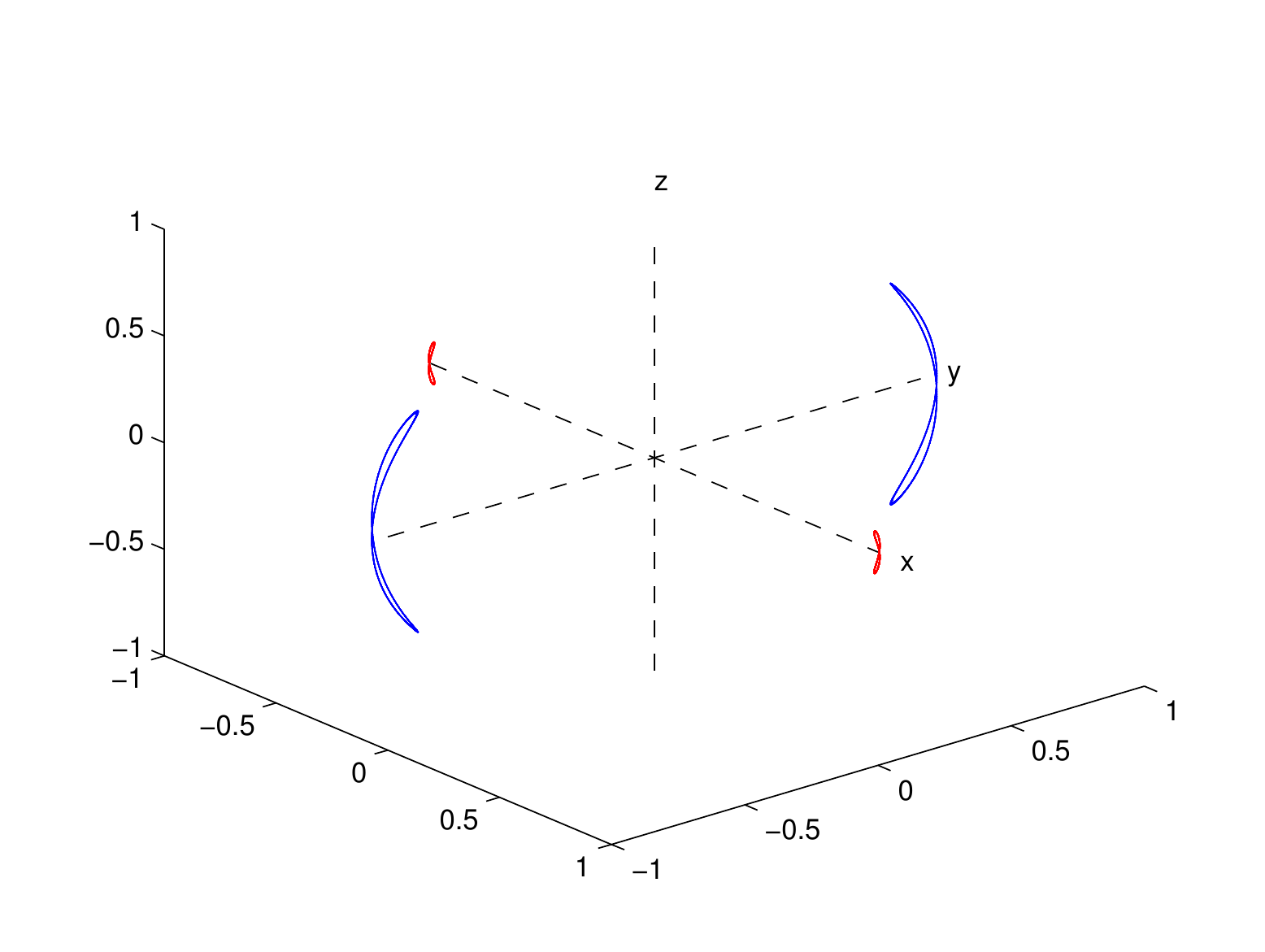}
  \caption{Left: Kayaking. The red and blue lines are motions of $\pm\bm{q}_1$ and $\pm\bm{q}_2$: when $\bm{q}_1$ corresponds to blue line, it shows K-$Q_1$ mode; otherwise it shows K-$Q_2$ mode. Right: Double splayed, where red line gives the motion of $\pm\bm{q}_2$, and blue line gives the motions of $\pm\bm{q}_1$. }
  \label{kay1}
\end{figure}

For the molecular model, the flow modes are described below. 
\begin{enumerate}
\item Log-rolling (LR): steady state, where $\bm{q}_2$ is along the 
$z$- (vortex) direction, and $\bm{q}_1$ lies in the $x$-$y$ (shear) plane. 
\item Kayaking(K): one of $\bm{q}_1$ and $\bm{q}_2$ rotates round the $z$-axis, while the other shows a splayed pattern (see Fig. \ref{kay1} left). If $\bm{q}_i$ rotates round the $z$-axis, we denote the flow mode as K-$Q_i$. 
\iffalse
\begin{figure}
  \centering
  \includegraphics[width=0.48\textwidth,keepaspectratio]{kay0.pdf}
  \caption{Kayaking. Red: $\bm{q}_2$; Blue: $\bm{q}_1$. }
  \label{kay0}
\end{figure}
\fi
\begin{figure}
  \centering
  \includegraphics[width=0.4\textwidth,keepaspectratio]{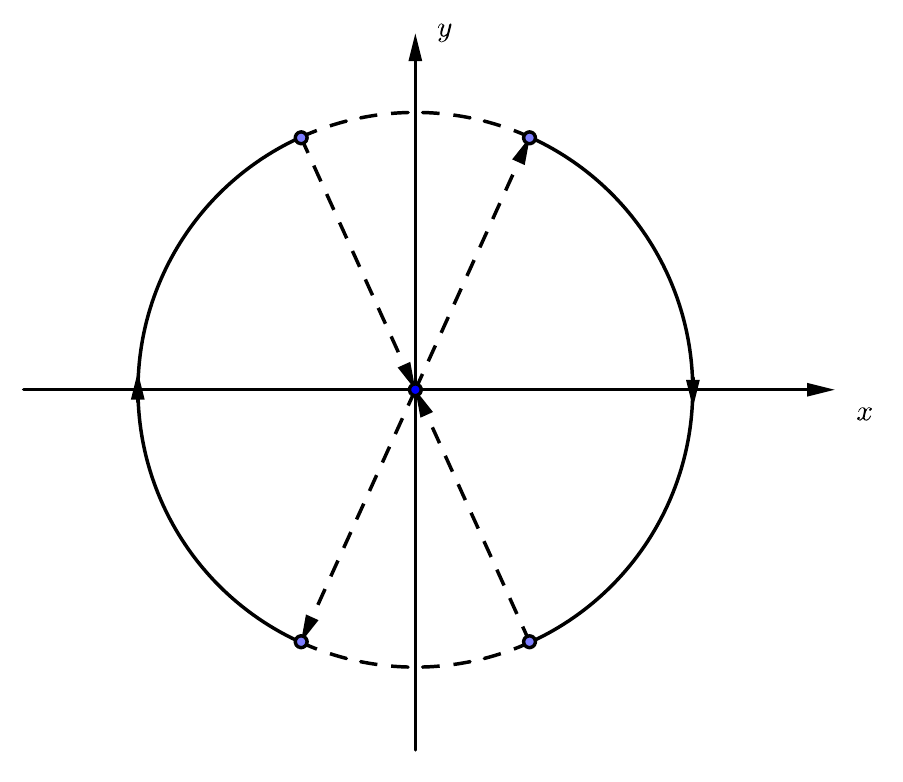}
  \includegraphics[width=0.4\textwidth,keepaspectratio]{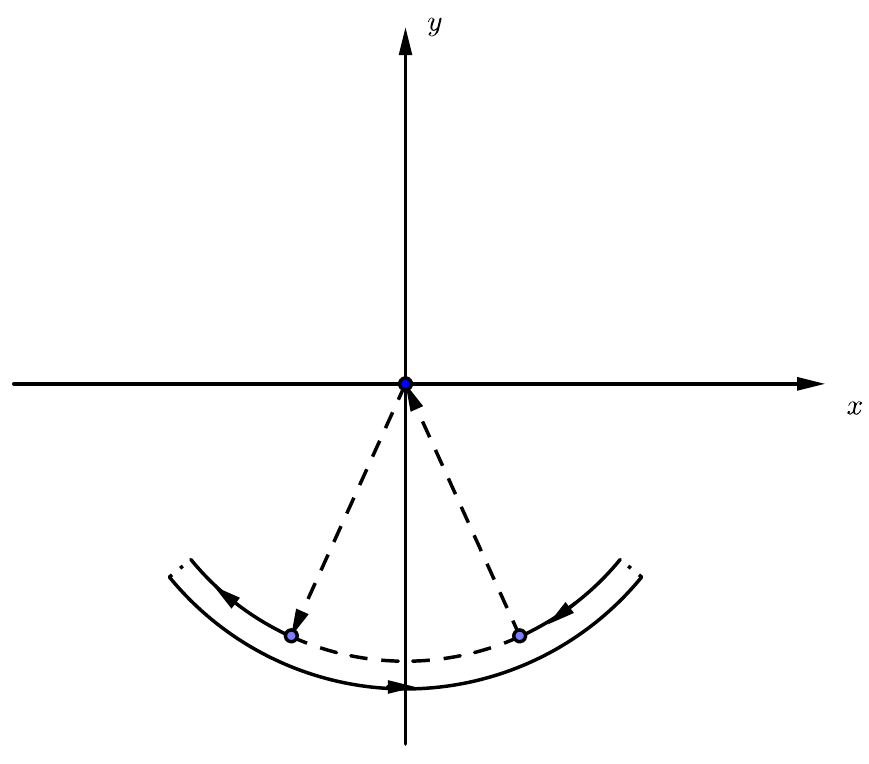}
  \caption{Left: Tumbling motion of $\bm{q}_1$. Right: Wagging-alternating motion of $\bm{q}_1$.}
  \label{tumb_wag}
\end{figure}
\item Double splayed (DS): $\bm{q}_2$ shows splayed pattern near the $x$-axis, $\bm{q}_1$ shows splayed near the $y$-axis (see Fig. \ref{kay1} right). 
\item Tumbling (T): $\bm{q}_2$ rotates in the $x$-$y$ plane; $\bm{q}_1$ also 
rotates in the plane, but jumps to $z$ when it approaches the $y$ axis (see Fig. \ref{tumb_wag} left). 
\item Wagging: $\bm{q}_2$ shows wagging near the $x$-axis. 
According to the motion of $\bm{q}_1$, it is further classified into two cases. 
  \begin{itemize}
  \item Wagging-alternating (W-A): $\bm{q}_1$ is wagging near the $y$-axis, with a jump to the $z$-axis (see Fig. \ref{tumb_wag} right). 
  \item Wagging-wagging (W-W): $\bm{q}_1$ is wagging near the $y$-axis. 
  \end{itemize}
\item Flow-aligning (FA): steady state, where $\bm{q}_2$ lies in the $x$-$y$ 
plane, while $\bm{q}_1$ may be in the $x$-$y$ plane (FA-$y$) or along the $z$ axis (FA-$z$). 
\end{enumerate}

\begin{table}
\begin{subtable}{\textwidth}
  \small\centering
  \begin{tabular}{|c|c|c|c|c|c|c|c|}
    \hline
    $\theta$ & LR & K-$Q_2$ & T & W-A & W-W & FA-$y$ & FA-$z$\\\hline
    $23\pi/32$ & - & $[0,10.0]$ & $[10.2,12.0]$ & $[12.2,14.8]$ & $[15.0,17.0]$ & $[17.2,20.0]$ & - \\\hline
    $22\pi/32$ & $[0.2,7.0]$ & $[7.2,8.6]$ & $[8.8,9.4]$ & $[9.6,12.0]$ & $[12.2,13.6]$ & $[13.8,20.0]$ & - \\\hline
    $21\pi/32$ & $[0.2,5.6]$ & - & $[5.8,6.0]$ & $[6.2,8.0]$ & $[8.2,9.6]$ & $[9.8,16.2]$ & $[16.4,20.0]$ \\\hline
    $20\pi/32$ & $[0.2,2.8]$ & $[3.0,3.4]$ & - & - & $[3.6,5.2]$ & $[5.4,11.0]$ & $[11.2,20.0]$ \\\hline
    $18\pi/32$ & - & - & - & - & - & - & $[0.2,20.0]$\\\hline
    $17\pi/32$ & - & - & - & - & - & - & $[0.2,20.0]$\\\hline
    $16\pi/32$ & - & - & - & - & - & - & $[0.2,20.0]$\\\hline
  \end{tabular}
  \caption{$\alpha=0.5$. }\label{phasemol1}
\end{subtable}\vspace{12pt}
\begin{subtable}{\textwidth}
  \small\centering
  \begin{tabular}{|c|c|c|c|c|c|c|c|}
    \hline
    $\theta$ & LR & K-$Q_2$ & T & W-A & W-W & FA-$y$ & FA-$z$\\\hline
    $23\pi/32$ & $[0.2,9.0]$ & $[9.2,9.4]$ & - & $[9.6,11.0]$ & $[11.2,12.0]$ & $[12.2,18.6]$ & $[18.8,20.0]$ \\\hline
    $22\pi/32$ & $[0.2,6.2]$ & $6.4$ & - & $[6.6,8.4]$ & $[8.6,9.2]$ & $[9.4,14.6]$ & $[14.8,20.0]$ \\\hline
    $21\pi/32$ & $[0.2,3.4]$ & $3.6$ & - & $[3.8,4.8]$ & $[5.0,5.8]$ & $[6.0,10.2]$ & $[10.4,20.0]$ \\\hline
    $20\pi/32$ & - & $[0.2,1.4]$ & - & $[1.6,2.0]$ & $[2.2,2.6]$ & $[2.8,6.0]$ & $[6.2,20.0]$ \\\hline
    $19\pi/32$ & - & - & - & - & - & - & $[0.2,20.0]$\\\hline
    $18\pi/32$ & - & - & - & - & - & - & $[0.2,20.0]$\\\hline
    $17\pi/32$ & - & - & - & - & - & - & $[0.2,20.0]$\\\hline
    $16\pi/32$ & - & - & - & - & - & - & $[0.2,20.0]$\\\hline
  \end{tabular}
  \caption{$\alpha=0.42$. }\label{phasemol2}
\end{subtable}
\caption{Range of the shear rate $k$ for flow modes in the molecular model for bent-core molecules. }
\end{table}
First we examine the flow modes for bent-core molecules. 
The range of shear rates for each flow mode in molecular model is listed in Table \ref{phasemol1} and Table \ref{phasemol2}. 
For $\theta=19\pi/32$ and $\alpha=0.5$, the flow modes are 
$$
\mbox{K-}Q_2:\ [0,2.8],\quad \mbox{K-}Q_1:\ [3.0,6.8],\quad \mbox{FA-}z: [7.0, 20.0]. 
$$
We have mentioned that in quiescent fluid, the bending angle $\theta=j\pi/32$ determines which of the three nematic phase is observed. 
In shear flow, different nematic phases result in distinct flow mode sequences with the shear rate $k$ increasing. 
When $16\le j\le 18$, namely the equilibrium phase is $N_3$, the only flow mode is FA-$z$. 
When $j=19$, namely the equilibrium phase is $B$, we obtain only FA-$z$ for $\alpha=0.42$, and the K-$Q_2$ --- K-$Q_1$ --- FA-$z$ sequence for $\alpha=0.5$. 
When $20\le j\le 23$, namely the equilibrium phase is $N_2$, the sequence follows LR --- K-$Q_2$ --- T --- W-A --- W-W --- FA-$y$ --- FA-$z$, with one or two modes missing. For $\alpha=0.42$, T is missing for all the four angles, and LR is not shown for $j=20$. For $\alpha=0.5$, LR is not found for $j=23$, K-$Q_2$ is not shown for $j=21$, T and W-A are missing for $j=20$, and FA-$z$ is not exhibited for $j=22,23$. 

\begin{table}
  \small\centering
  \begin{tabular}{|c|c|c|c|c|c|c|c|c|}
    \hline
    $l_2/l$ & LR & K-$Q_2$ & W-W & DS & FA-$y$ & K-$Q_1$ & FA-$z$\\\hline
    $0.125$ & $[0.2,4.4]$ & $[4.6,5.2]$ & $[5.4,6.8]$ & - & $[7.0,11.8]$ & - & $[12.0,20.0]$\\\hline
    $0.15$ & $[0.2,4.2]$ & $[4.4,5.0]$ & $[5.2,6.2]$ & - & $[6.4,11.0]$ & $[11.2,12.8]$ & $[13.0,20.0]$\\\hline
    $0.175$ & $[0.2,3.8]$ & $[4.0,4.4]$ & $[4.6, 4.8]$ & - & $[5.0,10.4]$ & $[10.6,12.4]$ & $[12.6,20.0]$ \\\hline
    $0.2$ & $[0.2,3.0]$ & $[10.0,10.6]$ & $[3.2,4.4]$ & - & $[4.6,9.8]$ & $[10.8,11.2]$ & $[11.4,20.0]$\\\hline
    $0.225$ & $[9.8,11.6]$ & $[0.2,1.8]$, $[9.2,9.6]$ & $[2.0,3.0]$ & - & $[3.2,9.0]$ & - & $[11.8,20.0]$\\\hline
    $0.25$ & $[8.8,12.6]$ & $[0.2,1.0]$ & $[1.2,1.8]$ & - & $[2.0,8.6]$ & - & $[12.8,20.0]$\\\hline
    $0.275$ & $[8.4,13.4]$ & - & - & $[0.2,2.4]$ & $[2.6,8.2]$ & - & $[13.6,20.0]$\\\hline
  \end{tabular}
  \caption{Range of the shear rate $k$ for flow modes in the molecular model for star molecules, $\alpha=0.42$. }\label{phase_star}
\end{table}
For star molecules, we can have a closer look at the effect of molecular shape in the biaxial region. 
As $l_2/l$ increases, the equilibrium phase sequence is $N_2$ --- $B$ --- $N_3$. 
At $l_2/l=0.125$, the flow mode sequence is LR --- K-$Q_2$ --- W-W --- FA-$y$ --- FA-$z$, which is part of the sequence found for bent-core molecules in the $N_2$ region. 
When $l_2/l$ increases and enters the $B$ region, the K-$Q_1$ mode emerges between FA-$y$ and FA-$z$. 
Then we observe some subtle phenomena when $l_2/l$ further increases. 
At $l_2/l=0.2$, the K-$Q_2$ mode moves to the middle of FA-$y$ and K-$Q_1$. 
At $l_2/l=0.225$, the mode at low shear rates changes from LR to K-$Q_2$ and LR emerges at high shear rates, resulting in the sequence K-$Q_2$ --- W-W --- FA-$y$ --- K-$Q_2$ --- LR --- FA-$z$. 
Then at $l_2/l=0.25$, K-$Q_2$ at high shear rates vanishes. 
Finally at $l_2/l=0.275$, the two periodic modes K-$Q_2$ and W-W are substituted by DS. 

Because rod-like molecules also show the $N_2$ phase in equilibrium, we would like to compare the flow modes of bent-core molecules with $\theta=j\pi/32,\, 20\le j\le 23$ with those of rod-like molecules that have been studied extensively in literature. If we only look at the motion of $\bm{q}_2$, the five modes are also found for rod-like molecules. 
The out-of-plane steady and out-of-plane oscillating states are also exhibited by rod-like molecules but are not shown in our results. 
The works on rod-like molecules imply that the occurrence of two out-of-plane modes might require a careful choice of $\alpha$ near the isotropic-nematic transition (see the solution diagrams in \cite{Dyn4}). Our choice of $\alpha$, however, is significantly larger than the transition value. 
%Besides, since we choose a fixed initial condition, we did not discuss the possibility of multiple flow modes under certain parameters. 

\subsubsection{Tensor model}
We only examine the bent-core molecules using the tensor model. 
The range of shear rates for each flow mode in tensor models is listed in Table \ref{phase1} and Table \ref{phase2}. 
For $\theta=19\pi/32$ and $\alpha=0.5$, the flow modes are 
$$
\mbox{DS}:\ [0,8.4],\quad \mbox{K-}Q_1:\ [8.6,16.8],\quad \mbox{FA-}z: [17.0, 20.0]. 
$$
We compare the results of tensor model with molecular model. 
Although under different shape parameters, we can observe all the flow modes found in the molecular model. 
For $16\le j\le 18$, the only mode FA-$z$ is the same as molecular model. 
This is also the case for $j=19$ and $\alpha=0.42$. 
For $j=19$ and $\alpha=0.5$, the DS takes the place of K-$Q_2$, while the K-$Q_1$ and FA-$z$ coincide with the molecular model. 
For $j=20$, the sequence in the tensor model covers that in the molecular model, with the extra LR for $\alpha=0.42$ and W-A for $\alpha=0.5$. 
For $21\le j\le 23$, the tensor model captures only the modes occurring at lower shear rate in the molecular model. 
Specifically, we do not observe W-W, FA-$y$ and FA-$z$ for the three $j$, and W-A for $j=23$ and $\alpha=0.5$. 
Some missing modes in molecular model are exhibited, including T for $\alpha=0.42$ and $j=22,23$, K-$Q_2$ for $\alpha=0.5$ and $j=21$, and LR for $\alpha=0.5$ and $j=23$. 

For the rod-like molecules, the tensor model with Bingham closure has been examined and compared with the molecular model. 
The results (see \cite{clos2,Yu2}) suggest that the tensor model works better when $\alpha$ is near the isotropic-nematic transition value, and at low shear rate. This is also observed in our results for bent-core molecules for $20\le j\le 23$. 
Because the I-N transition value increases as $\theta$ decreases, the $\alpha$ we choose is nearer to the I-N transition value for $j=20$ than $21\le j\le 23$. Indeed, for $j=20$, the flow mode sequence in tensor model better reflects that in molecular model. 

\begin{table}
\begin{subtable}{\textwidth}
  \small\centering
  \begin{tabular}{|c|c|c|c|c|c|c|c|c|}
    \hline
    $\theta$ & LR & K-$Q_2$ & T & W-A & W-W & FA-$y$ & FA-$z$\\\hline
    $23\pi/32$ & $[0.2,12.2]$ & $[12.4,19.2]$ & $[19.4,20.0]$ & - & - & - & -\\\hline
    $22\pi/32$ & $[0.2,8.6]$ & $[8.8,13.0]$ & $[13.2,14.4]$ & $[14.6,20.0]$ & - & - & -\\\hline
    $21\pi/32$ & $[0.2,4.8]$ & $[5.0,7.4]$ & $[7.6, 8.6]$ & $[8.8,20.0]$ & - & - & - \\\hline
    $20\pi/32$ & $[0.2,1.4]$ & $[1.6,3.0]$ & - & $[3.2, 5.8]$ & $[6.0,14.6]$ & $[14.8,19.6]$ & $[19.8,20.0]$\\\hline
    $18\pi/32$ & - & - & - & - & - & - & $[0.2,20.0]$\\\hline
    $17\pi/32$ & - & - & - & - & - & - & $[0.2,20.0]$\\\hline
    $16\pi/32$ & - & - & - & - & - & - & $[0.2,20.0]$\\\hline
  \end{tabular}
  \caption{$\alpha=0.5$. }\label{phase1}
\end{subtable}\vspace{12pt}
\begin{subtable}{\textwidth}
  \small\centering
  \begin{tabular}{|c|c|c|c|c|c|c|c|c|}
    \hline
    $\theta$ & LR & K-$Q_2$ & T & W-A & W-W & FA-$y$ & FA-$z$\\\hline
    $23\pi/32$ & $[0.2,7.4]$ & $[7.6,10.4]$ & $[10.6,12.0]$ & $[12.2,20.0]$ & - & - & -\\\hline
    $22\pi/32$ & $[0.2,4.4]$ & $[4.6,6.6]$ & $[6.8,7.4]$ & $[7.6,20.0]$ & - & - & -\\\hline
    $21\pi/32$ & $[0.2,1.8]$ & $[2.0,3.4]$ & - & $[3.6,20.0]$ & - & - & - \\\hline
    $20\pi/32$ & $[0.2,0.6]$ & $[0.8,1.0]$ & - & $[1.2, 1.8]$ & $[2.0,7.4]$ & $[7.6,10.4]$ & $[10.6,20.0]$\\\hline
    $19\pi/32$ & - & - & - & - & - & - & $[0.2,20.0]$\\\hline
    $18\pi/32$ & - & - & - & - & - & - & $[0.2,20.0]$\\\hline
    $17\pi/32$ & - & - & - & - & - & - & $[0.2,20.0]$\\\hline
    $16\pi/32$ & - & - & - & - & - & - & $[0.2,20.0]$\\\hline
  \end{tabular}
  \caption{$\alpha=0.42$. }\label{phase2}
\end{subtable}
\caption{Range of the shear rate $k$ for flow modes in the tensor model for bent-core molecules. }
\end{table}

\subsection{Discussion}
We compare our model and the results with those in \cite{cms2010}, where simulation is done for a molecular model, namely a model of the density function $f$. 

From the modeling viewpoint, the model in \cite{cms2010} has some limitations. 
It adopts a simple free energy, where the order parameters only include $Q_1$ and $Q_2$, which are not complete to describe the symmetry of the bent-core molecules. 
Also, the elastic energy and the spatial diffusion are also not included. 
Since bent-core molecules are able to show modulated nematic phases (see \cite{BentModel}), the model in \cite{cms2010} is only appropriate for homogeneous flows. 
Our model takes the above three aspects into account, which enables us to study inhomogeneous flows in the future. 

Another thing we would like to point out is that \cite{cms2010} derives the terms in the model using distinct molecular architectures, as stated in that work. 
This leads to different expressions from what we obtain of the rotational diffusion coefficients, the rotational convection, and the stress tensor. 
Usually this approach can be a good approximation, but is insufficient if we aim to investigate the effect of molecular architecture. 
Moreover, we note that the coefficients of the bulk free energy are different. 
%We plot in Fig. \ref{coe} the coefficients in our model and in \cite{cms2010}. 
The difference in coefficients may lead to different phase behavior, thus may cause significant distinction in flow modes, which is indeed reflected in our results. 
\iffalse
\begin{figure}
  \centering
  \includegraphics[width=0.4\textwidth,keepaspectratio]{par1.pdf}
  \caption{Bulk coefficients. Left: in our model; Right: in \cite{cms2010}.} 
  \label{coe}
\end{figure}
\fi

We observe some similar results in our simulation and in \cite{cms2010}. 
We confirm that $\bm{p}=0$, validating the choice of bulk energy without $\bm{p}$ in the homogeneous case. 
Also, the occurrence of LR phase at low shear rate and FA phase at high shear rate is identical. 
However, the periodic flow modes are different, 
which can be expected because there are many differences in the terms and coefficients. 
Our model shows some flow modes analogous to those of rod-like molecules, 
while \cite{cms2010} reported some modes where the motion of $\bm{q}_i$ are tilted. 

Finally, we discuss the mixed moments $\left<\bm{m}_i\bm{m}_j\right>,\,(i\ne j)$. 
In our model, they are zero in both molecular and tensor model, which is consistent with the molecular symmetry. 
We can also prove it directly from the dynamic model. 
Actually, in the tensor model, from the Boltzmann distribution, $\left<\bm{m}_i\bm{m}_j\right>=0$ for $i\ne j$. Also, in the molecular model, if the equality
$$
f(P(\alpha,\beta,\gamma),t)=f(P(\alpha,\beta,\gamma+\pi),t)=f(P(\pi-\alpha,\beta+\pi,\pi-\gamma),t)=f(P(\pi-\alpha,\beta,-\gamma),t) 
$$ 
holds for $t=0$, we can prove that it holds for arbitrary $t>0$ (see Appendix). 
We then derive from the above symmetric property that $\left<\bm{m}_i\bm{m}_j\right>=0$ for $i\ne j$. 

\section{Conclusion\label{concl}}
In this work, we establish the molecular model and tensor model for the 
dynamics of bent-core molecules and star molecules in incompressible fluid. 
We assume that the molecule is rigid consisting of spheres. 
Based on this architecture, we build hardcore molecular interaction 
and sphere--fluid friction into the model. 
In this way, we obtain the molecular model fully determined by physical parameters, which clearly reflects the molecular shape in the model. 
The molecular model incorporates three tensors describing polar and biaxial order, elastic energy that is able to describe the modulation, convection and diffusion both spatially and orientationally, and the corresponding stress and external force that let the system be energy dissipative. 
The tensor model is then derived from the molecular model. 
Along with the quasi-equilibrium closure approximation, the tensor model is also energy dissipative. 

We use both molecular model and tensor model to examine the flow modes in the shear flow problem. 
We focus on the effect of bending angle of bent-core molecules, and the length of the extra arm of star molecules. 
The parameters are chosen to cover the transition between three nematic phases $N_2$, $B$ and $N_3$. 
We examine the change of flow modes when the parameters go across the $N_2$ -- $B$ and $B$ -- $N_3$ phase boundaries. 
When the equilibrium phase is $N_2$, we find the flow modes analogous to the rod-like molecules. 
The tensor model is able to capture all the flow modes shown by the molecular model. 
Under our choice of parameters, the flow mode sequence is mostly identical to that in the molecular model for smaller bending angles, and recovers the part of sequence in the molecular model at low shear rate for larger bending angles. 

Although we only examined the shear flow problem, both the molecular model and the tensor model are ready for the studies of inhomogeneous flows. 
Also, for the shear flow problem, the model can be applied to other molecules with the same symmetry to carry out extensive investigations of the effect of molecular shape. 
The efficient implementation of closure approximation is also worth discussion. 
Currently we compute the quasi-equilibrium approximation by integration, which is time-consuming. 
Besides, the formulation of the tensor model allows us to adopt simpler closure approximations that might be suitable for specific types of flows, 
which is expected to be done in the future. 
%More simulations need to be done to understand the flow modes under different concentration and bending angle. 
%The formulation is not restricted to bent-core molecules, thus we can apply our model to other molecules, for instance, star molecules, as we have examined in \cite{BentModel}. 

\appendix
\section{Some equalities about the closure}
Suppose $f$ is given by (\ref{form1}), and $Z$ is defined in (\ref{PartFunc}). 
Direct computation gives
\begin{equation}
\frac{F_{entropy}}{k_BT}=\int\d\nu\rho(\bm{x},P)\log\rho(\bm{x},P)
=\bm{b}\cdot\bm{p}+B_1:Q_1+B_2:Q_2-\log Z, \label{Entropy1}
\end{equation}
and 
\begin{equation}
  \frac{\partial\log Z}{\partial (\bm{b},B_1,B_2)}=
  \frac{1}{Z}\frac{\partial Z}{\partial (\bm{b},B_1,B_2)}=(\bm{p},Q_1,Q_2). 
\end{equation}
Thus we can deduce that (see \cite{BentModel})
\begin{equation}\label{average}
  \frac{1}{k_BT}\frac{\partial F_{entropy}}{\partial (\bm{p},Q_1,Q_2)}=(\bm{b},B_1,B_2). 
\end{equation}
%
\iffalse
Take $\bm{p}$ as an example. We have 
\begin{align*}
  &\frac{1}{k_BT}\frac{\partial F_{entropy}}{\partial \bm{p}}\\
  =&\frac{\partial (\bm{b}\cdot\bm{p}+B_1:Q_1+B_2:Q_2-\log Z)}
  {\partial \bm{p}}\\
  =&\bm{b}
  +\bm{p}\cdot\frac{\partial \bm{b}}{\partial \bm{p}}
  +Q_1:\frac{\partial B_1}{\partial \bm{p}}
  +Q_2:\frac{\partial B_2}{\partial \bm{p}}
  -\frac{\partial \log Z}{\partial \bm{p}}\\
  =&\bm{b}
  +\frac{\partial \log Z}{\partial \bm{b}}\cdot\frac{\partial \bm{b}}{\partial \bm{p}}
  +\frac{\partial \log Z}{\partial B_1}:\frac{\partial B_1}{\partial \bm{p}}
  +\frac{\partial \log Z}{\partial B_2}:\frac{\partial B_2}{\partial \bm{p}}
  -\frac{\partial \log Z}{\partial \bm{p}}\\
  =&\bm{b}. 
\end{align*}
\fi
%
Moreover, The Jacobian $\frac{\partial (\bm{p},Q_1,Q_2)}{\partial (\bm{b},B_1,B_2)}$ 
can be expressed by high-order tensors, 
\begin{align}\label{covariance}
\frac{\partial (\bm{p},Q_1,Q_2)}{\partial (\bm{b},B_1,B_2)}=&
\frac{\partial^2 \log Z}{\partial (\bm{b},B_1,B_2)^2}\nonumber\\
=&\left<(\bm{m}_1,\bm{m}_1\bm{m}_1,\bm{m}_2\bm{m}_2)(\bm{m}_1,\bm{m}_1\bm{m}_1,\bm{m}_2\bm{m}_2)\right>-(\bm{p},Q_1,Q_2)(\bm{p},Q_1,Q_2)\nonumber\\
=&\mbox{cov}(\bm{m}_1,\bm{m}_1\bm{m}_1,\bm{m}_2\bm{m}_2). 
\end{align}
%
\iffalse
We give one example. 
\begin{align*}
  \frac{\partial Q_1}{\partial B_2}
  =&\frac{\partial}{\partial B_2}\left(\
  \frac{1}{Z}\frac{\partial Z}{\partial B_1}
  \right)\\
  =&\frac{1}{Z}\frac{\partial^2 Z}{\partial B_1\partial B_2}
  -\frac{1}{Z^2}\frac{\partial Z}{\partial B_1}\frac{\partial Z}{\partial B_2}\\
  =&\left<\bm{m}_1\bm{m}_1\bm{m}_2\bm{m}_2\right>-Q_1Q_2. 
\end{align*}
\fi

\section{The proof of the energy dissipation law of the tensor model}
Now we prove the energy dissipation law, 
for which we need to rewrite the diffusion terms. 
From (\ref{average}), we can write
\begin{equation}
  \mu=k_BT+\mu_{\bm{p}}\cdot \bm{m}_1+\mu_{Q_1}:\bm{m}_1\bm{m}_1+\mu_{Q_2}:\bm{m}_2\bm{m}_2.
\end{equation}
Here we denote $\mu_{\bm{p}}=\delta F/\delta \bm{p}$ etc., and $F$ is the free energy. 
Thus the terms like (\ref{ComputeRight}) are rewritten as 
\begin{align}
&-\int\d\nu \bm{m}_1\bm{m}_1D_2L_2^2f
=D_2\int\d\nu fL_2(\bm{m}_1\bm{m}_1)L_2(\log f)\nonumber\\
=&D_2\int\d\nu f(-\bm{m}_1\bm{m}_3-\bm{m}_3\bm{m}_1)
(-\bm{b}\cdot \bm{m}_3-B_1:(\bm{m}_1\bm{m}_3+\bm{m}_3\bm{m}_1))\nonumber\\
=&D_2\left[\bm{b}\cdot\left<\bm{m}_3(\bm{m}_3\bm{m}_1+\bm{m}_1\bm{m}_3)\right>
+B_1:\left<(\bm{m}_1\bm{m}_3+\bm{m}_3\bm{m}_1)(\bm{m}_1\bm{m}_3+\bm{m}_3\bm{m}_1)\right>\right]. \label{Right2}
\end{align}
Now we can rewrite 
\begin{align}
  -\mathcal{M}_{\bm{p}}=&D_2\left[\mu_{\bm{p}}\cdot\left<\bm{m}_3\bm{m}_3\right>
  +\mu_{Q_1}:(\left<\bm{m}_1\bm{m}_3\bm{m}_3\right>
  +\left<\bm{m}_3\bm{m}_1\bm{m}_3\right>)\right]\nonumber\\
  &+D_3\left[\mu_{\bm{p}}\cdot\left<\bm{m}_2\bm{m}_2\right>
  +(\mu_{Q_1}-\mu_{Q_2}):(\left<\bm{m}_1\bm{m}_2\bm{m}_2\right>
  +\left<\bm{m}_2\bm{m}_1\bm{m}_2\right>)\right].\nonumber\\
  %\\
  -\mathcal{M}_{Q_1}=&D_2\left[\mu_{\bm{p}}\cdot
  (\left<\bm{m}_3\bm{m}_3\bm{m}_1\right>
  +\left<\bm{m}_3\bm{m}_1\bm{m}_3\right>)\right.\nonumber\\
  &+\left.\mu_{Q_1}:\left<(\bm{m}_1\bm{m}_3+\bm{m}_3\bm{m}_1)
  (\bm{m}_1\bm{m}_3+\bm{m}_3\bm{m}_1)\right>\right]\nonumber\\
  &+D_3\left[\mu_{\bm{p}}\cdot
  (\left<\bm{m}_2\bm{m}_2\bm{m}_1\right>
  +\left<\bm{m}_2\bm{m}_1\bm{m}_2\right>)\right.\nonumber\\
  &\left.+(\mu_{Q_1}-\mu_{Q_2}):\left<(\bm{m}_1\bm{m}_2+\bm{m}_2\bm{m}_1)
  (\bm{m}_1\bm{m}_2+\bm{m}_2\bm{m}_1)\right>\right].\nonumber\\
  %\\
  -\mathcal{M}_{Q_2}=&D_1\mu_{Q_2}:\left<(\bm{m}_2\bm{m}_3+\bm{m}_3\bm{m}_2)
  (\bm{m}_2\bm{m}_3+\bm{m}_3\bm{m}_2)\right>
  \nonumber\\
  &+D_3\left[-\mu_{\bm{p}}\cdot
  (\left<\bm{m}_2\bm{m}_2\bm{m}_1\right>
  +\left<\bm{m}_2\bm{m}_1\bm{m}_2\right>)\right.\nonumber\\
  &\left.-(\mu_{Q_1}-\mu_{Q_2}):\left<(\bm{m}_1\bm{m}_2+\bm{m}_2\bm{m}_1)
  (\bm{m}_1\bm{m}_2+\bm{m}_2\bm{m}_1)\right>\right].\nonumber\\
  %\\
  \mathcal{V}_{\bm{p}}=&\kappa:\big[\left<\bm{m}_1\bm{m}_3\bm{m}_3\right>
  +\frac{I_{22}}{I_{11}+I_{22}}\left<\bm{m}_1\bm{m}_2\bm{m}_2\right>
  -\frac{I_{11}}{I_{11}+I_{22}}\left<\bm{m}_2\bm{m}_1\bm{m}_2\right>\big].\nonumber\\
  \left(\mathcal{N}_{\bm{p}}\right)_{\alpha}=&
  \partial_i\left(\partial_j(\mu_{\bm{p}})_k\sum_{\sigma=1}^3\gamma_{\sigma}
  \left<m_{1k}m_{1\alpha}m_{\sigma i}m_{\sigma j}\right>\right.\nonumber\\
  &+\partial_j(\mu_{Q_1})_{kl}\sum_{\sigma=1}^3\gamma_{\sigma}
  \left<m_{1k}m_{1l}m_{1\alpha}m_{\sigma i}m_{\sigma j}\right>
  \nonumber\\
  &\left.+\partial_j(\mu_{Q_2})_{kl}\sum_{\sigma=1}^3\gamma_{\sigma}
  \left<m_{2k}m_{2l}m_{1\alpha}m_{\sigma i}m_{\sigma j}\right>
  \right).\nonumber\\
  \left(\mathcal{N}_{Q_1}\right)_{\alpha\beta}=&
  \partial_i\left(\partial_j(\mu_{\bm{p}})_k\sum_{\sigma=1}^3\gamma_{\sigma}
  \left<m_{1k}m_{1\alpha}m_{1\beta}m_{\sigma i}m_{\sigma j}\right>\right.\nonumber\\
  &+\partial_j(\mu_{Q_1})_{kl}\sum_{\sigma=1}^3\gamma_{\sigma}
  \left<m_{1k}m_{1l}m_{1\alpha}m_{1\beta}m_{\sigma i}m_{\sigma j}\right>\nonumber\\
  &\left.+\partial_j(\mu_{Q_2})_{kl}\sum_{\sigma=1}^3\gamma_{\sigma}
  \left<m_{2k}m_{2l}m_{1\alpha}m_{1\beta}m_{\sigma i}m_{\sigma j}\right>\right).\nonumber\\
  \left(\mathcal{N}_{Q_2}\right)_{\alpha\beta}=&
  \partial_i\left(\partial_j(\mu_{\bm{p}})_k\sum_{\sigma=1}^3\gamma_{\sigma}
  \left<m_{1k}m_{2\alpha}m_{2\beta}m_{\sigma i}m_{\sigma j}\right>\right.\nonumber\\
  &+\partial_j(\mu_{Q_1})_{kl}\sum_{\sigma=1}^3\gamma_{\sigma}
  \left<m_{1k}m_{1l}m_{2\alpha}m_{2\beta}m_{\sigma i}m_{\sigma j}\right>\nonumber\\
  &\left.+\partial_j(\mu_{Q_2})_{kl}\sum_{\sigma=1}^3\gamma_{\sigma}
  \left<m_{2k}m_{2l}m_{2\alpha}m_{2\beta}m_{\sigma i}m_{\sigma j}\right>\right).\nonumber
\end{align}
Meanwhile, we rewrite the elastic stress as 
\begin{align}
  \tau_e=&ck_BT\big\{
  \mu_{Q_2}:\left<(\bm{m}_2\bm{m}_3+\bm{m}_3\bm{m}_2)\bm{m}_2\bm{m}_3\right>
  +\mu_{\bm{p}}\cdot\left<\bm{m}_3\bm{m}_1\bm{m}_3\right>\nonumber\\
  &+\mu_{Q_1}:\left<(\bm{m}_1\bm{m}_3+\bm{m}_3\bm{m}_1)\bm{m}_1\bm{m}_3\right>
  \nonumber\\
  &+\frac{1}{I_{11}+I_{22}}\big[\mu_{\bm{p}}\cdot\left<\bm{m}_2
    (I_{22}\bm{m}_1\bm{m}_2
    -I_{11}\bm{m}_2\bm{m}_1)\right>\nonumber\\
  &+(\mu_{Q_1}-\mu_{Q_2}):\left<(\bm{m}_1\bm{m}_2+\bm{m}_2\bm{m}_1)
    (I_{22}\bm{m}_1\bm{m}_2
    -I_{11}\bm{m}_2\bm{m}_1)\right>\big]\big\}.
\end{align}
From the above equations, we deduce (\ref{disp}). 

\section{Kirkwood theory}
We describe how to calculate the spatial diffusion coefficient matrix $\bm{J}$ using the Kirkwood theory. 
Assume that the molecule consists of $N$ spheres. 
Denote by $\bm{F}_i$ the force exerted on the sphere $i$ due to hydrodynamic interaction. 
The Kirkwood theory gives 
\begin{equation}
  \bm{V}_i=\sum_{j}H_{ij}\bm{F}_j, \label{Kirk}
\end{equation}
where 
\begin{equation}
  H_{ij}=\frac{1}{8\pi\eta_0|\hat{\bm{r}}_{ij}|}(I+\frac{\hat{\bm{r}}_{ij}\hat{\bm{r}}_{ij}}{|\hat{\bm{r}}_{ij}|^2}),\qquad 
  \hat{\bm{r}}_{ij}=\hat{\bm{r}}_j-\hat{\bm{r}}_i,\qquad j\ne i. 
\end{equation}
For $j=i$, we adopt the approximation $H_{ii}=I/\tau$ \cite{Doi_book}, where $I$ is the identity matrix. 
We choose $\tau=32\pi\eta_0 D$. 
Suppose that a molecule is undergoing a translation in the quiescent fluid with 
the velocity $\bm{V}$. Then $\bm{V}_i=\bm{V}$. 
On the other hand, the total hydrodynamic force shall be identical to the force 
that stems from the thermodynamic potential. Thus we have 
\begin{equation}
  \sum_{i}\bm{F}_i=-\nabla\mu. \label{force_bal}
\end{equation}
From (\ref{Kirk}) and (\ref{force_bal}), we can deduce the relation of $\bm{V}$ 
and $\nabla\mu$. 
Define $H\in\mathbb{R}^{3N\times 3N}$, $L\in\mathbb{R}^{3\times 3N}$ and $\bm{F}\in\mathbb{R}^{3N}$ by 
\begin{align*}
  H=\left(
  \begin{array}{ccc}
    H_{11} & \hdots & H_{1N}\\
    \vdots & & \vdots\\
    H_{N1} & \hdots & H_{NN}
  \end{array}
  \right), \quad
  L=(\underbrace{I,I,\ldots,I}_N), \quad
  \bm{F}=\left(
  \begin{array}{c}
    \bm{F}_1\\ \vdots \\ \bm{F}_N
  \end{array}
  \right). 
\end{align*}
Then we can rewrite \eqref{Kirk} and \eqref{force_bal} as
$$
H\bm{F}=L^T\bm{V},\quad L\bm{F}=-\nabla\mu. 
$$
Therefore, we can solve that 
$$
\bm{V}=-(LH^{-1}L^T)^{-1}\nabla\mu. 
$$
Thus, $\bm{J}=(LH^{-1}L^T)^{-1}$. 

For bent-core molecules, we use a discrete version of (\ref{sphdis}), namely to view the molecule as consisting of $1+N=1+1/\eta$ spheres located at
\begin{equation}
  \hat{\bm{r}}_j=l(\frac{1}{4}-|s_j|)\cos\frac{\theta}{2}\bm{m}_1
  +ls_j\sin\frac{\theta}{2}\bm{m}_2, \label{sphdis_dis}
\end{equation}
where $s_j=j/N,\ -N/2\le j\le N/2$. 
Using this molecular architecture we arrive at \eqref{Jcoef}. 
\iffalse
We find that the diffusion matrix $\bm{J}$ is diagonal, so it can be written as 
\begin{equation}
  \bm{J}=\frac{1}{8\pi\eta_sl}\sum_{j=1}^3\gamma_j^{-1}\bm{m}_j\bm{m}_j. 
\end{equation}
We plot $\gamma_j$ in Fig. \ref{difcoe}. 
\fi

\section{Symmetry of the molecular model in homogeneous case}
We investigate the Smochulowski equation in the shear flow, 
\begin{align}
\frac{\partial f(P,t)}{\partial t}
=L\cdot [(D_0\bm{I}^{-1})(k_BTLf+fLV)] - L\cdot(\bm{g}f), 
\label{eqn_f_h}
\end{align}
where $g$ is given by \eqref{grot_bent}, $V$ is given by \eqref{Vtensor} and \eqref{V_begin}-\eqref{V_end} without gradient terms. 
We will prove that if the equality
$$
f(P(\alpha,\beta,\gamma),t)=f(P(\alpha,\beta,\gamma+\pi),t)=f(P(\pi-\alpha,\beta+\pi,\pi-\gamma),t)=f(P(\pi-\alpha,\beta,-\gamma),t) 
$$ 
holds for $t=0$, then it holds for $t>0$. 

We only prove the first equality, because the other two follow exactly the same way. 
By \eqref{DiffRep}, for arbitrary $u$, we have 
\begin{align*}
(L_1u)(P(\alpha,\beta,\gamma),t)=&L_1\Big(u(P(\alpha,\beta,\gamma+\pi),t)\Big),\\
(L_2u)(P(\alpha,\beta,\gamma),t)=&-L_2\Big(u(P(\alpha,\beta,\gamma+\pi),t)\Big),\\
(L_3u)(P(\alpha,\beta,\gamma),t)=&-L_3\Big(u(P(\alpha,\beta,\gamma+\pi),t)\Big).
\end{align*}
We then examine the symmetry of right-hand terms at $t=0$. 
By the symmetry of $f$, we also have $V(P(\alpha,\beta,\gamma),0)=V(P(\alpha,\beta,\gamma+\pi),0)$. 
Thus, we can verify that for the diffusion term, 
\begin{align*}
&\Big[L\cdot \Big((D_0\bm{I}^{-1})(k_BTLf+fLV)\Big))\Big](P(\alpha,\beta,\gamma),0)\\&\quad=\Big[L\cdot \Big((D_0\bm{I}^{-1})(k_BTLf+fLV)\Big)\Big](P(\alpha,\beta,\gamma+\pi),0). 
\end{align*}
For the convection term, write $\bm{g}=\sum_{i=1}^3(\kappa:\alpha_i)\bm{m}_i$. 
It is straightforward to verify that 
\begin{align*}
  \alpha_1(P(\alpha,\beta,\gamma),0)=&\alpha_1(P(\alpha,\beta,\gamma+\pi),0), \\
  \alpha_2(P(\alpha,\beta,\gamma),0)=&-\alpha_2(P(\alpha,\beta,\gamma+\pi),0), \\
  \alpha_3(P(\alpha,\beta,\gamma),0)=&-\alpha_3(P(\alpha,\beta,\gamma+\pi),0). 
\end{align*}
Hence, 
$$
[L\cdot(\bm{g}f)](P(\alpha,\beta,\gamma),0)=[L\cdot(\bm{g}f)](P(\alpha,\beta,\gamma+\pi),0). 
$$
Therefore, $f(P(\alpha,\beta,\gamma),t)$ and $f(P(\alpha,\beta,\gamma+\pi),t)$ are governed by the same equation. Since they are equal at $t=0$, it is also the case for any $t>0$. 

\vspace{12pt}
\textbf{Acknowledgments.} Pingwen Zhang is partially supported by National Natural Science Foundation of China (Grant No. 11421101 and 11421110001). 

\bibliographystyle{plain}
\bibliography{bib_dyn}

\begin{thebibliography}{10}

\bibitem{BiExp_prl2004_2}
B.~R. Acharya, A.~Primak, and S.~Kumar.
\newblock Biaxial nematic phase in bent-core thermotropic mesogens.
\newblock {\em Phys. Rev. Lett.}, 92:145506, 2004.

\bibitem{advani1987use}
S.~G. Advani and C.~L. Tucker~III.
\newblock The use of tensors to describe and predict fiber orientation in short
  fiber composites.
\newblock {\em J. Rheol.}, 31(8):751--784, 1987.

\bibitem{advani1990closure}
S.~G. Advani and C.~L. Tucker~III.
\newblock Closure approximations for three-dimensional structure tensors.
\newblock {\em J. Rheol.}, 34(3):367--386, 1990.

\bibitem{Beris_book}
A.~N. Beris and B.~J. Edwards.
\newblock {\em {Thermodynamics of Flowing Systems with Internal
  Microstructure}}.
\newblock Oxford University Press, 1994.

\bibitem{pre73}
F.~Bisi, E.~G. Virga, and E.~C.~Gartland et~al.
\newblock Universal mean-field phase diagram for biaxial nematics obtained from
  a minimax principle.
\newblock {\em Phys. Rev. E}, 73:051709, 2006.

\bibitem{Ntb}
V.~Borshch, Y.-K. Kim, J.~Xiang, M.~Gao, A.~J\'akli, V.~P. Panov, J.~K. Vij,
  C.~T. Imrie, M.~G. Tamba, G.~H. Mehl, and O.~D. Lavrentovich.
\newblock Nematic twist-bend phase with nanoscale modulation of molecular
  orientation.
\newblock {\em Nat. Commun.}, 4:2635, 2013.

\bibitem{EC_pra2}
H.~Brand and H.~Pleiner.
\newblock Hydrodynamics of biaxial discotics.
\newblock {\em Phys. Rev. A}, 24:2777, 1981.

\bibitem{clos3}
C.~V. Chaubal and L.~G. Leal.
\newblock A closure approximation for liquid-crystalline polymer models based
  on parametric density estimation.
\newblock {\em J. Rheol.}, 42(1):177--201, 1998.

\bibitem{Ntb2}
D.~Chen, J.~H. Porada, J.~B. Hooper, A.~Klittnick, Y.~Shen, M.~R. Tuchband,
  E.~Korblova, D.~Bedrov, D.~M. Walba, M.~A. Glaser, J.~E. Maclennan, and N.~A.
  Clark.
\newblock Chiral heliconical ground state of nanoscale pitch in a nematic
  liquid crystal of achiral molecular dimers.
\newblock {\em Proc. Natl. Acad. Sci. USA}, 110:15931, 2013.

\bibitem{Doi_book}
M.~Doi and S.~F. Edwards.
\newblock {\em {The Theory of Polymer Dynamics}}.
\newblock Oxford University Press, 1986.

\bibitem{feng1998closure}
J.~Feng, C.~V. Chaubal, and L.~G. Leal.
\newblock {Closure approximations for the Doi theory: Which to use in
  simulating complex flows of liquid-crystalline polymers?}
\newblock {\em J. Rheol.}, 42(5):1095--1119, 1998.

\bibitem{Dyn4}
M.~G. Forest, Q.~Wang, and R.~Zhou.
\newblock {The flow-phase diagram of Doi--Hess theory for sheared nematic
  polymers II: finite share rates}.
\newblock {\em Rheol. Acta.}, 44(1):80–--93, 2004.

\bibitem{Dyn3}
M.~G. Forest, Q.~Wang, and R.~Zhou.
\newblock The weak shear kinetic phase diagram for nematic polymers.
\newblock {\em Rheol. Acta.}, 43(1):17–--37, 2004.

\bibitem{DynInh2}
M.~G. Forest, R.~Zhou, and Q.~Wang.
\newblock {Kinetic structure simulations of nematic polymers in plane Couette
  cells. II: in-plane structure transitions}.
\newblock {\em Multiscale Model. Simul.}, 4(4):1280--1304, 2005.

\bibitem{PRL_115_147805}
C.~Greco and A.~Ferrarini.
\newblock {Entropy-Driven Chiral Order in a System of Achiral Bent Particles}.
\newblock {\em Phys. Rev. Lett.}, 115:147805, 2015.

\bibitem{RodModel}
J.~Han, Y.~Luo, W.~Wang, P.~Zhang, and Z.~Zhang.
\newblock From microscopic theory to macroscopic theory: a systematic study on
  modeling for liquid crystals.
\newblock {\em Arch. Rat. Mech. Anal.}, 215:741, 2015.

\bibitem{clos1}
E.~J. Hinch and L.~G. Leal.
\newblock {Constitutive equations in suspension mechanics. Part II.
  Approximation forms for a suspension of rigid particles affected by Brownian
  rotations}.
\newblock {\em J. Fluid Mech.}, 76:187--208, 1976.

\bibitem{Ilg2}
P.~Ilg, I.~V. Karlin, {M. Kr\" oger}, and {H. C. \" Ottinger}.
\newblock {Canonical distribution functions in polymer dynamics. (II).
  Liquid-crystalline polymers}.
\newblock {\em Physica A}, 319:134--150, 2003.

\bibitem{clos2}
{J. S. Cintra, Jr.} and C.~L.~Tucker III.
\newblock Orthotropic closure approximations for flow-induced fiber
  orientation.
\newblock {\em J. Rheol.}, 39(6):1095, 1995.

\bibitem{Dyn1}
R.~G. Larson.
\newblock Arrested tumbling in shearing flows of liquid-crystal polymers.
\newblock {\em Macromolecules}, 23:3983--3992, 1990.

\bibitem{Dyn2}
R.~G. Larson and {H. C. \"Ottinger}.
\newblock Effect of molecular elasitcity on out-of-plane orientations in
  shearing flows of liquid-crystalline polymers.
\newblock {\em Macromolecules}, 24:6270--6282, 1991.

\bibitem{Ericksen_Leslie}
F.~M. Leslie.
\newblock Theory of flow phenomena in liquid crystals.
\newblock {\em Adv. Liq. Cryst.}, 4:1--81, 1979.

\bibitem{EC_pra1}
M.~Liu.
\newblock Hydrodynamic theory of biaxial nematics.
\newblock {\em Phys. Rev. A}, 24:2720, 1981.

\bibitem{BiExp_prl2004}
L.~A. Madsen, T.~J. Dingemans, M.~Nakata, and E.~T. Samulski.
\newblock Thermotropic biaxial nematic liquid crystals.
\newblock {\em Phys. Rev. Lett.}, 92:145505, 2004.

\bibitem{Marrucci_JNNFM1992}
G.~Marrucci and F.~Greco.
\newblock {A molecular approach to the polydomain structure of LCPs in weak
  shear flows}.
\newblock {\em J. Non-Newtonian Fluid Mech.}, 44:1--13, 1992.

\bibitem{prl_111_067801}
C.~Meyer, G.~R. Luckhurst, and I.~Dozov.
\newblock {Flexoelectrically Driven Electroclinic Effect in the Twist-Bend
  Nematic Phase of Achiral Molecules with Bent Shapes}.
\newblock {\em Phys. Rev. Lett.}, 111:067801, 2013.

\bibitem{pre1998}
T.~Qian and P.~Sheng.
\newblock Generalized hydrodynamic equations for nematic liquid crystals.
\newblock {\em Phys. Rev. E}, 58:7475--7485, 1998.

\bibitem{EC_pra3}
W.~M. Saslow.
\newblock Hydrodynamics of biaxial nematics with arbitrary nonsingular
  textures.
\newblock {\em Phys. Rev. A}, 25:3350, 1982.

\bibitem{prl_3D_2014}
S.~M. Shamid, D.~W. Allender, and J.~V. Selinger.
\newblock {Predicting a Polar Analog of Chiral Blue Phases in Liquid Crystals}.
\newblock {\em Phys. Rev. Lett.}, 113:237801, 2014.

\bibitem{pre2013}
S.~M. Shamid, S.~Dhakal, and J.~V. Selinger.
\newblock Statistical mechanics of bend flexoelectricity and the twist-bend
  phase in bent-core liquid crystals.
\newblock {\em Phys. Rev. E}, 87:052503, 2013.

\bibitem{cms2010}
S.~Sircar, J.~Li, and Q.~Wang.
\newblock Biaxial phases of bent-core liquid crystal polymers in shear flows.
\newblock {\em Comm. Math. Sci.}, 8:697--720, 2010.

\bibitem{pre78}
S.~Sircar and Q.~Wang.
\newblock Shear-induced mesostructures in biaxial liquid crystals.
\newblock {\em Phys. Rev. E}, 78:061702, 2008.

\bibitem{jr2009}
S.~Sircar and Q.~Wang.
\newblock Dynamics and rheology of biaxial liquid crystal polymers in shear
  flow.
\newblock {\em J. Rheol.}, 53:819--858, 2009.

\bibitem{pre2014_2}
E.~G. Virga.
\newblock Double-well elastic theory for twist-bend nematic phases.
\newblock {\em Phys. Rev. E}, 89:052502, 2014.

\bibitem{wang1997comparative}
Q.~Wang.
\newblock {Comparative studies on closure approximations in flows of liquid
  crystal polymers: I. elongational flows}.
\newblock {\em J. Non-Newtonian Fluid Mech.}, 72(2-3):141--162, 1997.

\bibitem{WangELiuZhang_PRE2002}
Q.~Wang, W.~E, C.~Liu, and P.~Zhang.
\newblock Kinetic theory for flows of nonhomogeneous rodlike liquid crystalline
  polymers with a nonlocal intermolecular potential.
\newblock {\em Phys. Rev. E}, 65:051504, 2002.

\bibitem{Wigner_book}
E.~P. Wigner.
\newblock {\em {Group Theory and its Application to the Quantum Mechanics of
  Atomic Spectra}}.
\newblock {Academic Press, New York}, 1959.

\bibitem{BentModel}
J.~Xu, F.~Ye, and P.~Zhang.
\newblock A tensor model for nematic phases of bent-core molecules based on
  molecular theory.
\newblock {\em Submitted to Multiscale Model. Simul.}, arXiv:1408:3722v2, 2016.

\bibitem{SymmO}
J.~Xu and P.~Zhang.
\newblock From microscopic theory to macroscopic theory --- symmetries and
  order parameters of rigid molecules.
\newblock {\em Sci. China Math.}, 57:443--468, 2014.

\bibitem{Yu2}
H.~Yu, G.~Ji, and P.~Zhang.
\newblock {A Nonhomogeneous Kinetic Model of Liquid Crystal Polymers and Its
  Thermodynamic Closure Approximation}.
\newblock {\em Commun. Comput. Phys.}, 7(2):383--402, 2010.

\bibitem{Yu1}
H.~Yu and P.~Zhang.
\newblock A kinetic–hydrodynamic simulation of microstructure of liquid
  crystal polymers in plane shear flow.
\newblock {\em J. Non-Newtonian Fluid Mech.}, 141:116--127, 2007.

\bibitem{DynInh1}
R.~Zhou, M.~G. Forest, and Q.~Wang.
\newblock {Kinetic structure simulations of nematic polymers in plane couette
  cells. I: the algorithm and benchmarks}.
\newblock {\em Multiscale Model. Simul.}, 3(4):853--870, 2005.

\end{thebibliography}

\end{document}